\documentclass[twocolumn,pre,floatfix,superscriptaddress,letterpaper,nofootinbib]{revtex4-1}
\usepackage{hyperref}
\usepackage{amsmath,amssymb,amsfonts}
\usepackage{algorithmic}
\usepackage{graphicx}
\usepackage{textcomp}
\usepackage{xcolor}
\usepackage{soul}
\usepackage[braket, qm]{qcircuit}
\usepackage{braket}
\usepackage{multirow}
\usepackage{makecell}
\usepackage{booktabs}
\usepackage{array}
\usepackage{multirow}
\usepackage{wrapfig}
\usepackage{float}
\usepackage{colortbl}
\usepackage{tabu}
\usepackage{threeparttable}
\usepackage{threeparttablex}
\usepackage[normalem]{ulem}
\usepackage{subcaption}
\usepackage{orcidlink}
\usepackage{import}
\usepackage{datetime}
\usepackage{tikz}
\usepackage{mathtools}
\usepackage{listings}
\usepackage{url}

\newcounter{daggerfootnote}

\newif\ifdraft
\drafttrue
\ifdraft
\newcommand{\note}[1]{ {\textcolor{orange} { **: #1 }}}
\newcommand{\alnote}[1]{ {\textcolor{red} { ***Andre: #1 }}}
\newcommand{\crnote}[1]{ {\textcolor{green} { ***Carlos: #1 }}}
\newcommand{\jknote}[1]{ {\textcolor{blue} { ***Johannes: #1 }}}
\newcommand{\tenote}[1]{ {\textcolor{purple} { ***ThEh: #1 }}}
\newcommand{\fknote}[1]{ {\textcolor{pink} { ***Florian: #1 }}}
\newcommand{\cjnote}[1]{ {\textcolor{purple} { ***Caitlin: #1 }}}
\newcommand{\tsnote}[1]{ {\textcolor{magenta} { ***ThomasS: #1 }}}
\newcommand{\avnote}[1]{ {\textcolor{cyan} { ***Aleksandar: #1 }}}
\newcommand{\omnote}[1]{ {\textcolor{orange} { ***Oliver: #1 }}}
\newcommand{\jdnote}[1]{ {\textcolor{purple} { ***Joseph: #1 }}}
\else
\newcommand{\note}[1]{}
\newcommand{\alnote}[1]{}
\newcommand{\crnote}[1]{}
\newcommand{\jknote}[1]{}
\newcommand{\tenote}[1]{}
\newcommand{\etnote}[1]{}
\newcommand{\mpnote}[1]{}
\newcommand{\cjnote}[1]{}
\newcommand{\tsnote}[1]{}
\newcommand{\avnote}[1]{}
\newcommand{\omnote}[1]{}
\newcommand{\jdnote}[1]{}
\newcommand{\fknote}[1]{}
\fi

\begin{document}

\title{A performance characterization of quantum generative models
}

\author{Carlos~A.~Riofr\'io$^{\orcidlink{0000-0002-7346-9198}}$}
\affiliation{BMW Group, Munich, Germany}
\affiliation{QUTAC, Quantum Technology and Application Consortium, Germany}

\author{Oliver~Mitevski$^{\orcidlink{0000-0001-5949-3856}}$}
\affiliation{Munich Re, Munich, Germany}
\affiliation{QUTAC, Quantum Technology and Application Consortium, Germany}

\author{Caitlin Jones$^{\orcidlink{0000-0002-3317-9786}}$}
\affiliation{BASF Digital Solutions GmbH, Ludwigshafen am Rhein, Germany}
\affiliation{QUTAC, Quantum Technology and Application Consortium, Germany}

\author{Florian~Krellner$^{\orcidlink{0000-0001-9532-1656}}$}
\affiliation{SAP SE, Walldorf, Germany}
\affiliation{QUTAC, Quantum Technology and Application Consortium, Germany}

\author{Aleksandar~Vu\v{c}kovi\'{c}$^{\orcidlink{0000-0003-1329-2156}}$}
\affiliation{Faculty of Physics, TU Darmstadt, Darmstadt, Germany}
\affiliation{Healthcare R\&D Digital Innovation, Merck KGaA, Darmstadt, Germany}
\affiliation{QUTAC, Quantum Technology and Application Consortium, Germany}

\author{Joseph~Doetsch$^{\orcidlink{0000-0002-2927-9557}}$}
\affiliation{Lufthansa Industry Solutions, Norderstedt, Germany}
\affiliation{QUTAC, Quantum Technology and Application Consortium, Germany}

\author{Johannes~Klepsch$^{\orcidlink{0000-0002-8247-9590}}$}
\affiliation{BMW Group, Munich, Germany}
\affiliation{QUTAC, Quantum Technology and Application Consortium, Germany}

\author{Thomas~Ehmer$^{\orcidlink{0000-0002-4586-5361}}$}
\affiliation{Healthcare R\&D Digital Innovation, Merck KGaA, Darmstadt, Germany}
\affiliation{QUTAC, Quantum Technology and Application Consortium, Germany}

\author{Andre~Luckow$^{\orcidlink{0000-0002-1225-4062}}$}
\affiliation{BMW Group, Munich, Germany}
\affiliation{QUTAC, Quantum Technology and Application Consortium, Germany}
\date{\today}

%\thispagestyle{plain}
%\pagestyle{plain}

%\tableofcontents
%\import{}{abstract.tex}

\begin{abstract}

Quantum generative modeling is a growing area of interest for industry-relevant applications. This work systematically compares a broad range of techniques to guide quantum computing practitioners when deciding which models and methods to use in their applications. We compare fundamentally different architectural ansatzes of parametric quantum circuits: 1.\,A \emph{continuous} architecture, which produces continuous-valued data samples, and 2.\,a \emph{discrete} architecture, which samples on a discrete grid. We also compare the performance of different data transformations: the min-max and the probability integral transforms. We use two popular training methods: 1.\, quantum circuit Born machines (QCBM), and 2.\, quantum generative adversarial networks (QGAN). We study their performance and trade-offs as the number of model parameters increases, with a baseline comparison of similarly trained classical neural networks. The study is performed on six low-dimensional synthetic and two real financial data sets. Our two key findings are that: 1.\, For all data sets, our quantum models require similar or fewer parameters than their classical counterparts. In the extreme case, the quantum models require two orders of magnitude less parameters. 2.\, We empirically find that a variant of the \emph{discrete} architecture, which learns the copula of the probability distribution, outperforms all other methods.
\end{abstract}

%\collaboration{This paper was developed within the Quantum Technology and Application Consortium (QUTAC)}    
\maketitle

% Sections
\section{Introduction}

Quantum technologies are transitioning from a research topic to industrial applications. New quantum hardware architectures, with more qubits and less noise, and possible advantageous applications keep materializing in academic and industrial settings. Quantum computing devices currently do not surpass classical counterparts in practical applications, but rapid advancements suggest examination of near-term applications should be pursued. As an industry consortium, QUTAC\footnote{Quantum Technology \& Application Consortium, \href{https://www.qutac.de/?lang=en}{qutac.de}}, encompassing twelve of the largest German companies, we are interested in understanding the value that quantum technologies can bring, in the short and long term, to our respective industries. Our long-term efforts are focused on identifying the use cases for which quantum computing has the potential to bring a quantum advantage. 

Since the invention of classical generative adversarial networks (GANs) in 2014~\cite{NIPS2014_5ca3e9b1}, generative modeling has become an ever-growing area of research with applications across all industries. 
The list of applications is extensive including: 1.\,generating statistics of rare events or particular data patterns of interest (e.\,g., to improve statistical techniques and AI models for anomaly detection, fraud identification, market simulations, quality assurance, autonomous agents, recommender and vision systems, among others); 2.\,building generative design pipelines (e.\,g., for drug discovery, personalized assistants, food design, image generation, metaverse room design, text and language generation, among others); and 3.\,addressing data privacy concerns for data sharing through synthesized realistic data replacements (e.g., to accelerate innovation through the ``open sourcing of realistic data'', or to help companies on their journey-to-cloud with otherwise sensitive data)~\cite{10.1115/1.4053859}. 

Classical GANs have been extraordinarily successful in addressing the above applications. However, they still have many practical challenges, such as requiring large data sets, tuning many hyper-parameters, driving high-outcome variability, long model convergence times, and extensive computational resources needed for training. These challenges present an opportunity for quantum computing \cite{creswell2018}. Quantum models could lead to faster training, more expressive power, \cite{sim2019expressibility, du2020expressive, glasser2019expressive}, better or faster inference, or smaller energy footprint in practice \cite{Villalonga_2020}. 

Since quantum systems are inherently probabilistic, quantum states are a natural choice to model probability distributions. This was taken advantage of in the so-called \emph{supremacy} experiments, \cite{10.1145/1993636.1993682, ZHU2022240, 48651}, where random quantum circuits implemented in quantum computers were used to sample from classically intractable probability distributions. For industrial applications, however, the quantum circuits and probability distributions we seek to generate are not arbitrary but determined by the application. In this vein, quantum machine learning could offer a framework for learning quantum states for arbitrary and practically useful distributions better than their classical counterparts, e.\,g., \cite{PhysRevX.12.021037, Sweke2021quantumversus, Perdomo_Ortiz_2018}. These distributions can be either the final product, or a means to an end for another quantum algorithm that requires these distributions \cite{PhysRevResearch.4.023136}.

This work investigates quantum generative modeling, which promises potential near-term advantages on noisy intermediate-scale quantum (NISQ) hardware~\cite{preskill2018}. In particular, we have identified and characterized two fundamentally different quantum circuit architectures, which can be used as quantum generative models. We differentiate between \emph{discrete} and \emph{continuous} architectures. The former learns and produces samples from a discretized sampling space, while the latter can produce continuous-valued samples. Moreover, as described by Zhu et al. \cite{Zhu_2022}, quantum information theory allows for an efficient representation of the copula \cite{nelsen2007introduction} of arbitrary probability distributions, which has the potential to simplify the learning process of complicated probability distributions. That is why, in our tests, the \emph{discrete} architectures are additionally divided into \emph{copula} and \emph{standard} variants.

\begin{figure*}[t]
    \centering
    \includegraphics[width=1\textwidth]{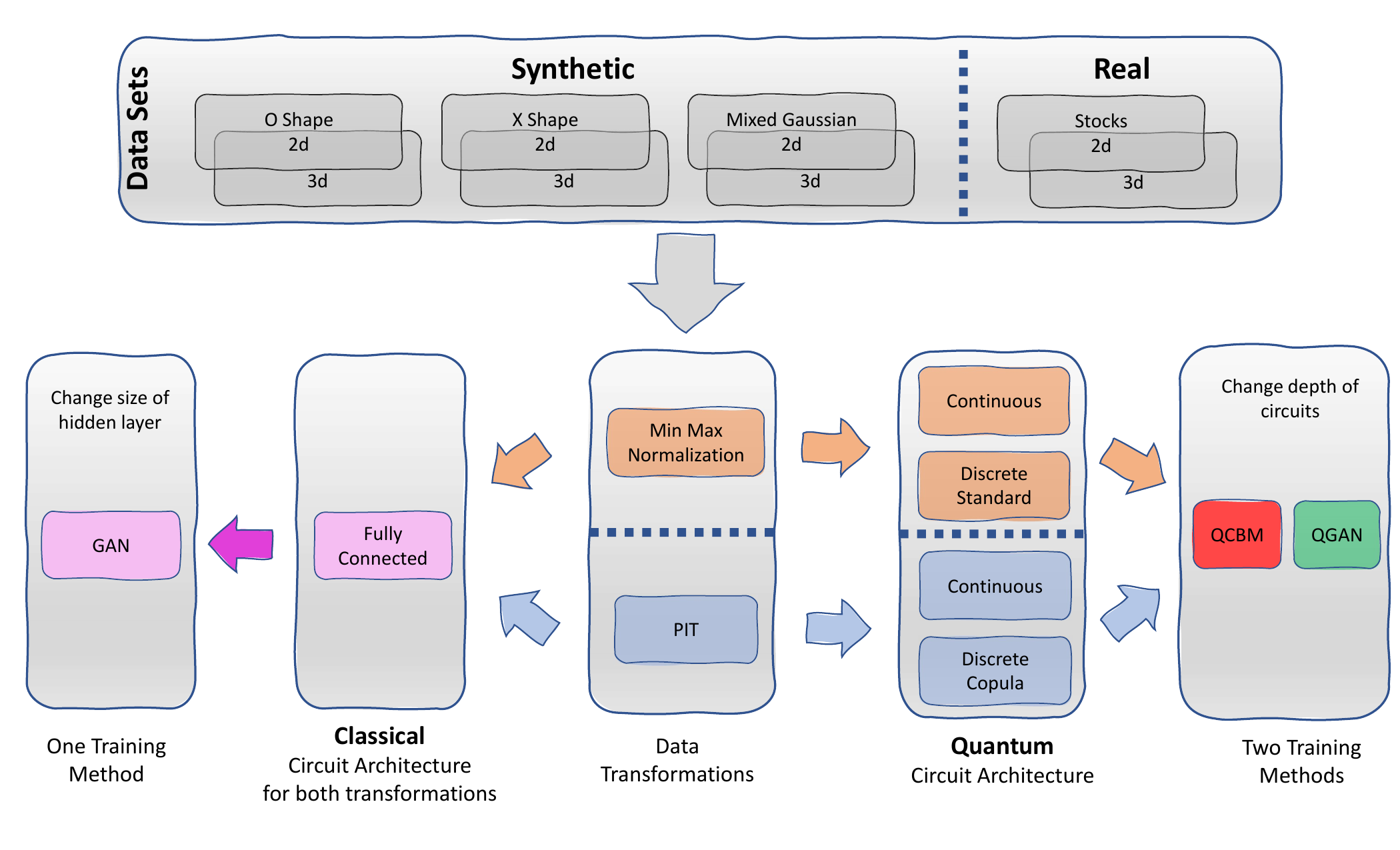}
    \caption{\textbf{Structure of the experiments}. Each experiment has a different data set, transformation, circuit architecture  and training method.  We benchmark the performance of each technique over different numbers of parameters, which is achieved by changing the size of the hidden layers in the classical case (left) and changing the depth of the circuit in the quantum case (right).
     }\label{experiment_structure}
\end{figure*}

In our evaluation, we consider two data transformations to pre-process the data before the learning starts. We investigate the differences between simple min-max normalization versus probability integral transform of the data, which is naturally matched to the \emph{discrete copula} architecture but can also be used with others. We use two common quantum generative approaches to train our models: quantum circuit Born machines (QCBMs) \cite{benedetti2019generative} and quantum GANs (QGANs). Both quantum generative techniques utilize hybrid quantum-classical approaches to machine learning \cite{schuld2015introduction, preskill2018,Perdomo_Ortiz_2018,Benedetti_2019}, which are suitable for the NISQ era of quantum computing. We show the structure of the experiments we study in this work in Fig.~\ref{experiment_structure}.

QCBMs have been shown to have great expressive power \cite{cheng2018information,gao2018,liu2018,zhu2019,Alcazar_2020,coyle2020born, Benedetti2021variational,zhu2022copula,DBLP:journals/corr/abs-2201-08770} and have been implemented on existing quantum hardware \cite{hamilton2019,zhu2019,leyton2021robust,rudolph2022generation}, including practical applications in finance \cite{Coyle_2021}. QGANs are the subject of much current study, with implementations which use a quantum generator and classical discriminator \cite{Situ_2020, Zeng2019learning, Zoufal2019quantum, Borras_Chang_2022impact, anand2021noise,BravoPrieto2022stylebasedquantum, zhouhybrid2022, Chang2022running} and those where both the generator and discriminator are quantum \cite{PhysRevA.98.012324, huang2021experimental, Niu2022entangling, chaudhary2022towards, 9605352}. 
Of these \cite{anand2021noise,Chang2022running, 9605352, Borras_Chang_2022impact}, also investigate the impact of noise on the performance of QGANs and \cite{Chang2022running, huang2021experimental, anand2021noise} deploy QGANs on quantum hardware.

For training the QCBMs, the approach we follow is to encode the histogram of training data sets directly into a quantum state. For the QGAN training, we use a hybrid approach with a quantum circuit for the generator and classical neural networks for the discriminator. All the optimization steps are classical updates of the variational parameters of the quantum circuits. We benchmark the training performance of all the architectures and training methods mentioned above using six synthetic and two real two- and three-dimensional data sets. We also train a simple classical neural-network-based generative model. As it is common practice in the field, we report the Kullback-Leibler divergence between the probability density of the training data and that of the generated samples as our performance measure. 

We compare the performance of the quantum models to  purely classical ones while varying the circuit depth, continuous or discrete architecture, and the data pre-processing. In a total, we ran of over 1200 numerical experiments and comparisons. We use the Qiskit and PennyLane state-vector simulators with Python for all experiments. We have open-sourced our simulation library, \href{https://github.com/QutacQuantum/qugen}{\emph{qugen}}, for reproducibility of the results. Our comparisons are intended to help users of quantum technologies decide which type of circuit architecture and data transformation to use for their applications. Our results highlight a trade-off between the size of the quantum register (number of qubits) and the time for training and performing inference with the generative model. The discrete models use more qubits and are faster to train and sample, whereas the continuous models are smaller, but training and sampling will take substantially longer in quantum hardware. Our results draw attention to the performance differences and opportunities between discrete and continuous architectures and different data transformation schemes. In fact, we show that the copula architecture, paired with the probability integral transform, has an edge in terms of training performance regarding the other quantum architectures tested in this work. For most data sets, the discrete copula architecture trained via the QCBM or QGAN is the most performant model we tested.

The rest of the paper is organized as follows: in Section \ref{sec:methodology}, we define all the architectures, data transformations, and training procedures we use for our benchmarks. In Section \ref{sec:results}, we present extensive numerical results of training QGANs and QCBMs using specially designed low-dimensional synthetic and real financial data. We show how different data transformations and architectures perform as we scale the size of the quantum circuits and the dimension of the data sets.
All our experiments were designed to be realistically deployed on existing quantum hardware. We end with a discussion about possible future work in Section \ref{sec:discussion}.

\section{Methodology}\label{sec:methodology}
In this section, we briefly review the concepts of generative modeling, starting with classical generative adversarial networks, and followed by two quantum training approaches: a QCBM and QGAN. 

\subsection{Generative modeling}
Generative models enable the creation of new data by learning the underlying probability distribution of the training data set of interest. More specifically, a generative model attempts to learn an unknown probability distribution $Q$  by modeling an approximated probability distribution $P(\boldsymbol{\theta})$ which is parameterized by a set of variables $\boldsymbol{\theta}$.  Data sampled from $Q$ is used to train a model by tuning $\boldsymbol{\theta}$ such that $P(\boldsymbol{\theta}$) more closely approximates $Q$. This process can be done in various ways, which we explain in the following sections.

\subsection{Classical generative adversarial networks}
\begin{figure}[t]
    \centering
    \includegraphics[width=0.5\textwidth]{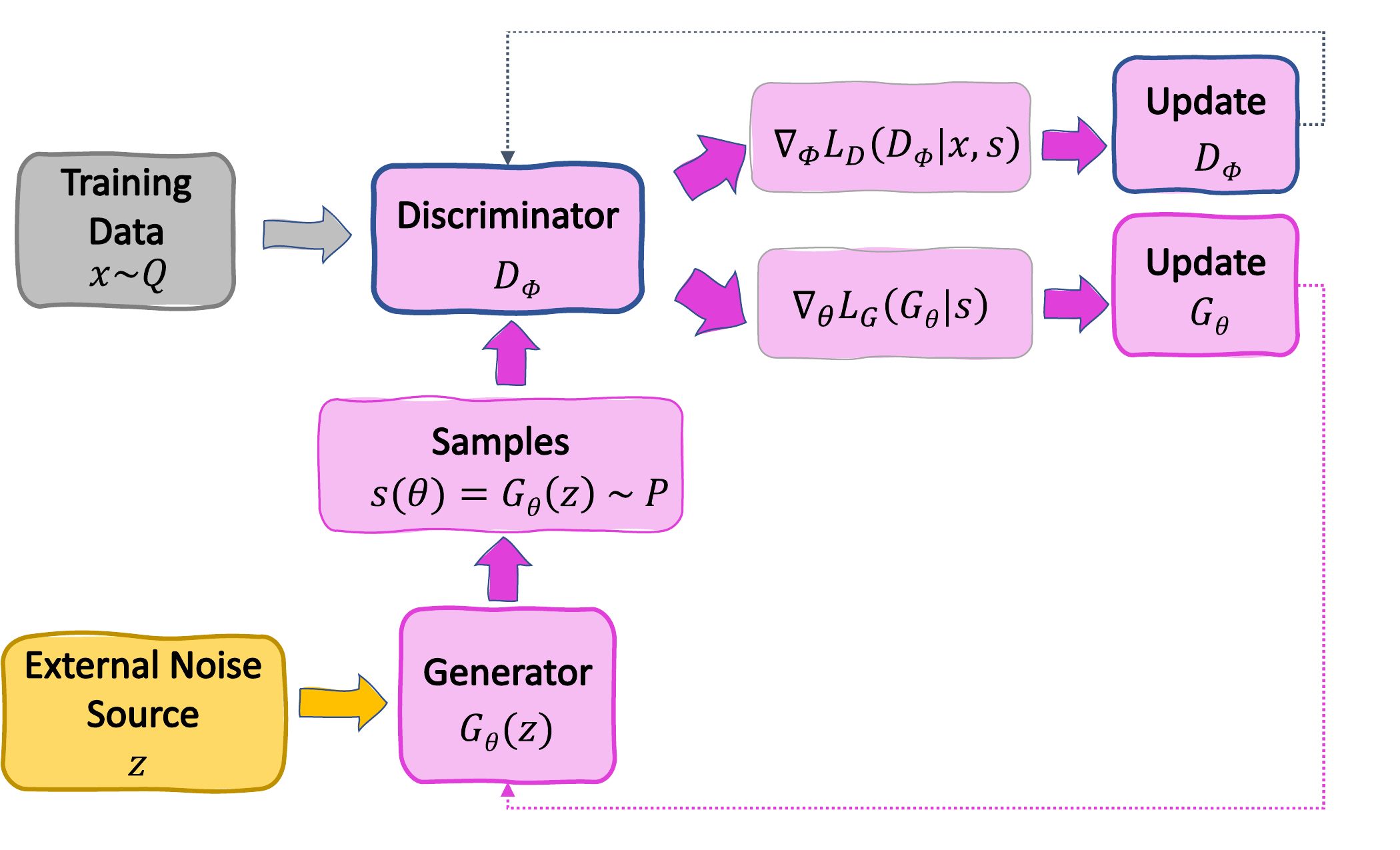}
    \caption{Schematic of classical GAN training. The generator $G_\theta$ learns how to produce samples, $s$ that are indistinguishable from real samples $x$. These are both fed to the discriminator $D$, which learns to distinguish real from generated samples. Note that this architecture needs an external source of noise (z) to generate synthetic samples.}\label{gan_training}
\end{figure}

One of the most widespread techniques for generative models is the Generative Adversarial Networks (GANs)~\cite{NIPS2014_5ca3e9b1}. Here, two neural networks, the discriminator~$D_{\boldsymbol{\phi}}$ and generator~$G_{\boldsymbol{\theta}}$, compete to learn the structure of the training data.  The generator attempts to fool the discriminator by generating data which is indistinguishable from the real training data and the discriminator learns to distinguish data created by the generator from the training data. A schematic representation of a typical GAN training loop is shown in Fig.~\ref{gan_training}.

As a result of this adversarial game, at the end of the training, the generator is capable of producing new data which behaves similarly to the real data. This approach, however, presents many challenges. For example, as these models are intrinsically hard to train, they usually require enormous amounts of data and complex network architectures, and they usually need large computational resources for training \cite{creswell2018}.

Mathematically, a GAN can be expressed as follows. A generator function $G_{\boldsymbol{\theta}}$ takes as an input a random variable (usually normally distributed) and generates a sample:
\begin{equation}
    G_{\boldsymbol{\theta}}(z_l) = s_l(\boldsymbol{\theta})
\end{equation}
where $z_l$ is an element of an $m$ dimensional vector $\mathbf{z} =\{z_1,...,z_l,...,z_m\}$ of random numbers  and $s_l$ is the corresponding sample created. Here, $m$ samples are generated from $m$ random numbers, which defines the batch size. 
A discriminator function $D_{\boldsymbol{\phi}}$ takes as input either a sample $s_l(\boldsymbol{\theta})$ generated by the generator or an element of the training data set $x_l$ which is an element of $\mathbf{x} =\{x_1,...,x_l,...,x_m\} , $ so the sampled and data batches both have $m$ data points. Each $x_l$ is sampled from the underlying distribution, $x_l \sim Q$. The discriminator function returns 1 if it classifies the sample as real and 0 if it deems it fake. A perfect discriminator would return $D_{\boldsymbol{\phi}}(x_l)= 1$ for all $x_l$ and  $D_{\boldsymbol{\phi}}(s_l(\boldsymbol{\theta}))= 0$ for all $s_l(\boldsymbol{\theta})$. 
The adversarial model is trained by alternating between minimizing the loss function for the generator, $\min_{\boldsymbol{\theta}} L_G(G_{\boldsymbol{\theta}})$, and the discriminator, $\min_{\boldsymbol{\phi}} L_D(D_{\boldsymbol{\phi}})$. Typically, the loss functions used for GANs are
\begin{multline}\label{eq:discriminator_loss}
     L_D(D_{\boldsymbol{\phi}}|\mathbf{x},\mathbf{s}) \\= -\frac{1}{m}\sum_{l=1}^m\left[\log(D_{\boldsymbol{\phi}}(x_l))+\log(1-D_{\boldsymbol{\phi}}(s_l(\boldsymbol{\theta})))\right],
\end{multline}
for the discriminator, and
\begin{equation}\label{eq:generator_loss}
     L_G(D_{\boldsymbol{\phi}} |\mathbf{s}) = -\frac{1}{m}\sum_{l=1}^m\log(D_{\boldsymbol{\phi}}(s_l(\boldsymbol{\theta}))),
\end{equation}
for the generator.

GANs are usually trained using gradient descent in the parameter space. The model parameters are updated proportionally to the gradient of the respective loss functions, Eqs.~\eqref{eq:discriminator_loss} and \eqref{eq:generator_loss}, given by $\nabla_{\boldsymbol{\phi}}L_D(D_{\boldsymbol{\phi}}|\textbf{x},\textbf{s})$ and $\nabla_{\boldsymbol{\theta}}L_G(G_{\boldsymbol{\theta}}|\textbf{s})$, to search for the minimum of both loss functions. The optimization is run for a set number of epochs (iterations) with a performance metric calculated each time. Then the model with the best fit to the empirical distribution $Q$ is chosen as the trained model. The performance metrics used in this work are discussed in section \ref{sec:metrics}. 

\subsection{Quantum circuit Born machine (QCBM)}\label{sec:QCBM}

QBCMs attempt to directly learn the probability distribution and do not have a discriminator. The task is to learn a unitary operator $\mathcal{U}(\boldsymbol{\theta})$ that maps an initial state, usually $|0\rangle_n = |0\rangle^{\otimes n}$, prepared in an $n$-qubit system, to some target state which encodes the probability distribution we want to learn
\begin{equation}\label{eq:state_preparation}
    |\psi\rangle = \mathcal{U}(\boldsymbol{\theta})|0\rangle_n.
\end{equation}
Here, $\boldsymbol{\theta}$ represents a set of parameters or degrees of freedom, which we must vary to learn the correct mapping. Depending on the level of access to quantum hardware, $\boldsymbol{\theta}$ can be low-level pulse control sequences or angles of quantum gates of a predetermined variational quantum circuit architecture. The nature of the target state $|\psi\rangle$ depends on the type of circuit architecture used. In quantum machine learning, one usually considers a variational circuit configuration depending on several interconnected parametric quantum gates; see for example \cite{Benedetti_2019}. The task is to find $\boldsymbol{\theta}$ so that we can approximate $|\psi\rangle$ as well as possible, given the available degrees of freedom.

In general, we are attempting to find a unitary $\mathcal{U}(\boldsymbol{\theta})$ to create a generator state $|G_{\boldsymbol{\theta}}\rangle$   
\begin{equation}\label{eq:generator_quantum_state}
    |G_{\boldsymbol{\theta}}\rangle =\mathcal{U}(\boldsymbol{\theta})|0\rangle_n = \sum_{i=0}^{2^n-1}c_i(\boldsymbol{\theta})|i\rangle ,
\end{equation}
where $|i\rangle$ are the computational basis elements for an $n$-qubit system, i.e., $i$ is written in its $n$-bit binary representation, $\{|00\ldots 0\rangle, |00\ldots 1\rangle, \ldots, |11\ldots 1\rangle\}$ and $c_i\in\mathbb{C}$ are the probability amplitudes determined by the value of $\boldsymbol{\theta}$. Thus, by repeated measurements of $| G_{\boldsymbol{\theta}} \rangle$ in the computational basis, the probability distribution $P(\boldsymbol{\theta})$ can be estimated. The exact method of encoding depends on the architecture of the quantum circuit, see Sec. \ref{subsec:circuit_arch} for explicit details of how this is achieved.

The learning problem is solved when a set of parameters $\boldsymbol{\theta}^*$ is found such that:
\begin{equation}\label{eq:kl_optimization}
    \boldsymbol{\theta}^* = \arg\min_{\boldsymbol{\theta}} C_{KL}(Q, P(\boldsymbol{\theta})),
\end{equation}
where $C_{KL}(Q, P(\boldsymbol{\theta}))$ is the KL divergence for two discretized probability distributions~$P$ and~$Q$:
\begin{equation}\label{eq:KL_loss}
    C_{KL}(Q, P) = \sum_{i=0}^{2^n-1}Q_i\ln{(Q_i/P_i)}
\end{equation}
and $Q$ and $P$ are described by two normalized vectors with entries $Q_i$ and $P_i$, respectively, usually computed as the histogram of the data samples.

\begin{figure}[t] 
    \centering
    \includegraphics[width=0.5\textwidth]{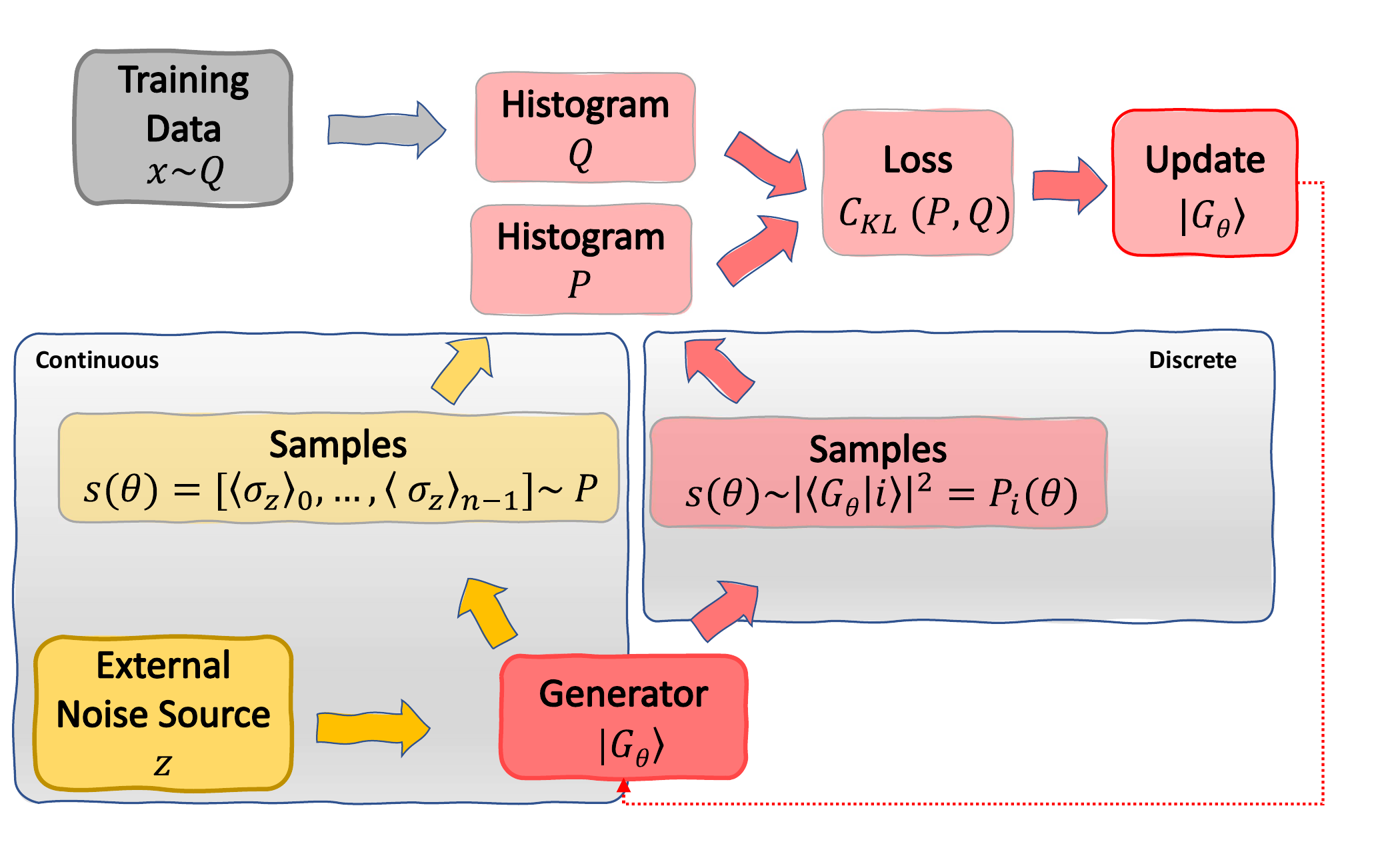}
    \caption{\textbf{Schematic of the training of a QCBM.}  A generator creates a generator state $| G_\theta  \rangle $ which when measured produces samples. The histogram of these samples $P$ can then be compared to the histogram from the training data $Q$ using the Kullback-Leibler divergence, $C_{KL}$. Note that for the continuous case (left path) there is an external source of noise to generate synthetic samples, whereas for the discrete case (right path) all randomness comes from the quantum measurement process.}\label{qcbm_training_combined}
\end{figure}

The optimization problem defined in Eq.~(\ref{eq:kl_optimization}) is handled classically, see for example \cite{schuld2021machine}. Hence, this type of approach is often termed hybrid quantum computing. A schematic showing a typical training loop for a QCBM is presented in Fig.~\ref{qcbm_training_combined}. For this model, the training is completed when the maximum allowed number of epochs (iterations) is reached or  when the change in the KL divergence between successive steps drops below a certain value, whichever comes first. The QCBM approach enables us to understand the limitations of the variational quantum circuit ansatz we have chosen, without the complications of the adversarial training of a GAN. The adversarial training itself is intricate and difficult to debug, as one needs to make sure that there is a balance between the discriminator and generator so that one model does not overpower the other. In practice, it is difficult to be certain whether poor performance is due to the circuit architecture or poor choice of learning rates, optimizer, or some other hyper-parameters. By solving a simpler optimization problem, we attempt to disentangle those effects to try to understand whether the variational quantum circuit architecture can, in principle, generate the desired state. 

\subsection{Quantum generative adversarial networks (QGAN)}\label{sec:QGAN}

Quantum generative adversarial modeling is the natural extension of generative modeling using quantum computing. A QGAN can be thought of as having the same structure as a conventional GAN except with either the generator or the discriminator (or both) replaced with a quantum circuit. In this work, we consider only the case where the generator is quantum, to reduce the number of qubits required. However, there is nothing in principle that prevents the discriminator from being realized in a fully quantum manner as done in \cite{PhysRevA.98.012324, huang2021experimental, Niu2022entangling, chaudhary2022towards, 9605352}.

\begin{figure}[t] 
    \centering
    \includegraphics[width=0.5\textwidth]{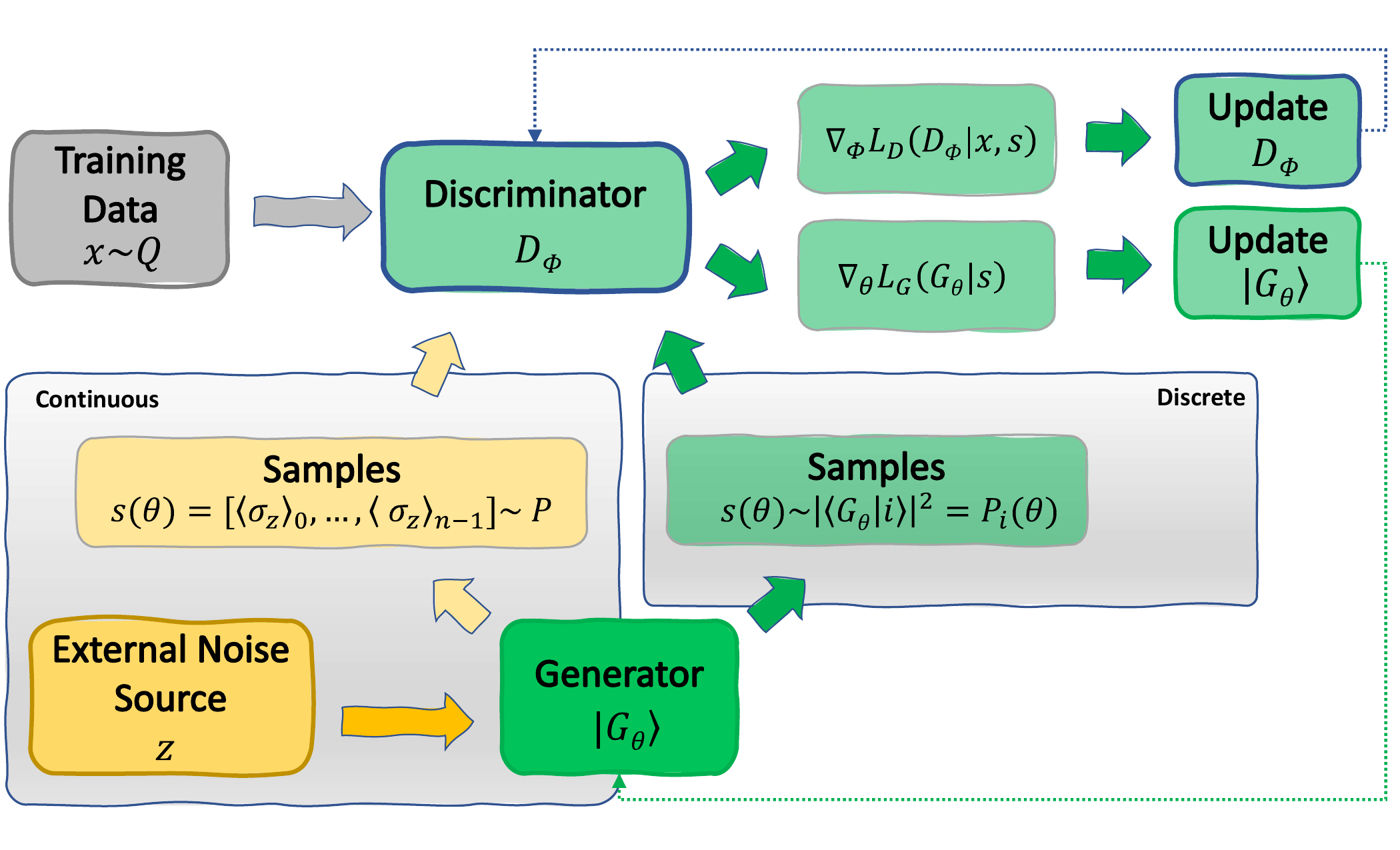}
    \caption{\textbf{Schematic of the training of a QGAN.} The main difference to the classical case is that the generator $G_{\boldsymbol{\theta}}$ is replaced by a quantum state $\ket{G_{\boldsymbol{\theta}}}$. For a continuous architecture (left path), an external source of noise (z) is needed to create the synthetic data samples, for the discrete architecture (right path), a single shot of the quantum circuit is equivalent to sampling a single data point and there is no need of an external noise source to create randomness.}\label{QGAN_training_combined}
\end{figure}

The generator is prepared in the same way as in the QCBM case, Eq.~(\ref{eq:generator_quantum_state}). This generator state can be used to learn the approximated probability distribution $P(\boldsymbol{\theta})$,  architecture-dependent details are given in subsection \ref{subsec:circuit_arch}. $P(\boldsymbol{\theta})$  is used to create samples of data $\mathbf{s}(\boldsymbol{\theta})$, which are then handed to a classical discriminator function that attempts to distinguish them from data $\mathbf{x}$ drawn from the underlying distribution $Q$.  As in the classical case, the loss function for both the discriminator and  generator are  given by Eqs.~(\ref{eq:discriminator_loss}) and (\ref{eq:generator_loss}), respectively. Then, alternating between the generator and discriminator, gradient descent is used to update the model during training. Again, classical hardware handles this optimization. We present a schematic of a typical training loop of a QGAN in Fig.~\ref{QGAN_training_combined}.

\subsection{Data transformations}\label{subsec:data_transforms}
Follow standard machine learning best practices, before running our benchmarking experiments, we need to pre-process the given data. We use and compare two different data transformations:
\subsubsection{Min-max normalization} In min-max normalization each data point is shifted such that it lies between $0$ and $1$:
\begin{equation}
    x'=\frac{x-x_{\min}}{x_{\max} - x_{\min}},
\end{equation}
where $x'$ is the normalized data set and $x_{\min}$ and $x_{\max}$ are the minimum and maximum values of the original data $x$, respectively.

\subsubsection{Probability integral transform (PIT)} 

For a random vector-valued variable $X: \Omega \rightarrow \mathbb{R}^d$ with a continuous distribution $Pr(X_1,\ldots,X_d)$ and cumulative marginal distribution function $F_i(X_i)$, defined as

\begin{equation}\label{eq:copula_marginals}
    \begin{split}
        F_i(x) & =Pr[X_i\leq x]\\
               & = \int_{-\infty}^{x}dx_i' \int_{\mathbb{R}}dx'_1 \cdots \int_{\mathbb{R}}dx'_d Pr(x_1',\ldots, x_i',\ldots, x_d'),
    \end{split}
\end{equation}
it is known that the random variable $U_i = F_i(X_i)$ is uniformly distributed in the interval $[0,1]$. As it is customary in the literature, we use upper case letters to denote a random variable and their corresponding lower case to denote a particular realization. 

The transformation of the original probability distribution to the distribution of its uniformly distributed cumulative marginals is known as the \emph{copula}. For details, see \cite{pit_angus}. Analytically, we can write the copula via the change-of-variable relation between probability densities as  

\begin{equation}\label{eq:copula_analytical}
\begin{split}
    & Pr(u_1,\ldots, u_d) = \\
    &\int_{\mathbb{R}} dx_1\cdots\int_{\mathbb{R}} dx_d \: \delta(U_1-u_1)\cdots\delta(U_d-u_d)Pr(x_1,\ldots, x_d) \\
    & =\int_{\mathbb{R}} dx_1\cdots\int_{\mathbb{R}} dx_d \: \delta(F_1(x_1)-u_1)\cdots\delta(F_d(x_d)-u_d) \times  \\ &Pr(x_1,\ldots, x_d),
\end{split}
\end{equation}
where $\delta(u)$ is the Dirac delta function \cite{van1992stochastic}. This equation computes the probability density function of the new variables $u_1,\ldots,u_d$. The PIT can also be approximately computed directly from data samples, which is how we use it in this work. In this case, one simply computes the histogram of the marginal distribution of the samples and finds the value of the random variable $U_i$ using Eq.~\eqref{eq:copula_marginals} substituting the probability distribution by the sampled histogram and the integrals by sums. When all $U_i$ variables are found, they are automatically distributed according to Eq.~\eqref{eq:copula_analytical}.

This transformation has the purpose of simplifying the structure of the data, as now all variables are correlated and uniformly distributed, independently of the nature of the original probability distribution. Under this transformation, the generative model learns the \emph{copula} of the joint probability distribution. In fact, it is known that learning the copula is equivalent to knowing the whole distribution, see for example \cite{nelsen2007introduction} for an introduction. For completeness, we show in the appendix \ref{appendix:pit_calculation}, an explicit calculation of the PIT for one of the data sets used in this work.

\subsection{Circuit architectures}\label{subsec:circuit_arch} We have identified two different quantum circuit ansatzes commonly used in the literature: 1.~A \emph{continuous}  architecture, resembling a classical neural network, which interprets the expectation values of the observables as the model output and directly produces continuous samples of the learned probability distribution architecture. This is also known as a \emph{deterministic} architecture \cite{schuld2021machine}. 2.~A \emph{discrete} architecture, which learns a discretized version of the probability distribution  and, in return, samples from it in a discrete fashion by interpreting the measurement outcomes as the model output. This is also known as \emph{probabilistic} architecture in \cite{schuld2021machine}. 

\subsubsection{Continuous}
This type of architecture is capable of learning and providing samples of probability distributions continuously. This architecture has been mainly explored in the supervised learning setting, see for example \cite{Mari2020transferlearning, Mitarai_2018}. Here we apply the same architecture for generative modeling, a practical example of which can be found in \cite{patel_cern_generative_2023}. 

In this case, the quantum computer learns a state
\begin{equation}\label{eq:generator_quantum_state_continuous}
    |G_{\boldsymbol{\theta}} (\boldsymbol{z})\rangle = \mathcal{U}(\mathbf{z},\boldsymbol{\theta})|0\rangle_n = \sum_{i=0}^{2^n-1}c_i(\mathbf{z},\boldsymbol{\theta})|i\rangle,
\end{equation}
where $\boldsymbol{\theta}$, as before, are the variational parameters of the quantum circuit, and $z$ represents a vector whose entries are randomly sampled, usually from the uniform or normal distributions, in the interval $[-\pi, \pi]$. Note that this architecture requires an external source of entropy to produce samples, analogous to the classical generative models. The noise is fed to the circuit via single qubit rotations (usually rotations around the $x$ axis) applied at the beginning of the quantum circuit. In our experiments, however, we adopt an architecture, sometimes referred to as \emph{data re-uploading} \cite{PerezSalinas2020datareuploading, PhysRevA.103.032430}, in which the noise is re-inserted in the circuit after each block of entangling and parametric gates is applied. We find this choice to work well in practice.

The output of this type of generative model is computed by estimating the expectation value of the Pauli Z operator, $\sigma_z$, in each qubit, after the circuit is applied, which requires many repetitions of the measurements or \emph{shots}.  Note that each noise realization $z$ produces a single data sample. 

The output of the model, for one sample, is given by the feature vector

\begin{equation}\label{eq:continuous_output_samples}
    s(\boldsymbol{\theta},\mathbf{z}) = [\langle \sigma_z (\boldsymbol{\theta},\mathbf{z})\rangle_0, \langle \sigma_z (\boldsymbol{\theta},\mathbf{z}) \rangle_1,\ldots,\langle \sigma_z (\boldsymbol{\theta},\mathbf{z})\rangle_{n-1}],
\end{equation}
where
\begin{multline}
\langle \sigma_z (\boldsymbol{\theta},\mathbf{z}) \rangle_k 
 =  \\\langle G_{\boldsymbol{\theta}} (\boldsymbol{z})|I_0\otimes\cdots\otimes I_{k-1}\otimes \sigma_{z}^{(k)} \otimes I_{k+1}\otimes\cdots\otimes I_{n-1} |G_{\boldsymbol{\theta}} (\boldsymbol{z})\rangle 
\end{multline} 
Since $\langle \sigma_z\rangle_k\in[-1,1]$, we usually normalize the samples via $(\langle \sigma_z\rangle_k + 1)/2$ so that they are in the interval $[0,1]$, which is more convenient to handle.

\begin{figure*}
\centering
\includegraphics[scale=0.8]{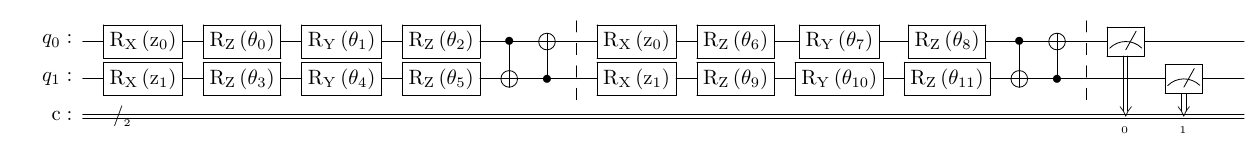}
\caption{\textbf{Continuous Circuit (2D data, two blocks).} This is the continuous circuit structure used for both the QCBM and QGAN generator.  Each block consists of eight rotation gates, six parameterized and two which take the random noise vector $z$, and two CNOT gates. $R_X(\theta)$ is a rotation of $\theta$ about the x axis. After the circuit is executed, many measurements in the computational basis are required to estimate the expectation values given in Eq.~\eqref{eq:continuous_output_samples}, and thus, to compute one sample of the generative model.}\label{fig:2D_continuous_circuit}
\end{figure*}

The main advantage of this architecture is that it can directly produce a continuously sampled vector as output. This is in contrast to the \emph{discrete} architecture described below. Another advantage is that it uses one qubit per dimension of the data, which is particularly convenient as current quantum hardware does not provide  more than a handful of qubits.  Its main drawback, however, is that it requires many measurement shots to construct the continuous output. 
This implies, in general, longer training and sample times. In this work, we used a state-vector simulator to compute the exact expectation values and do not rely on sampling, i.e., by collecting multiple measurement results and averaging them. This approach significantly reduced computation times. For an example of the continuous architecture used in this work, see Fig.~\ref{fig:2D_continuous_circuit}. This variational ansatz is inspired by \cite{PerezSalinas2020datareuploading, PhysRevA.103.032430}, in which the circuit inputs the noise via rotations followed by a series of single qubit rotations parametrized by the variational parameters $\mathbf{\theta}$, and a series of 2-qubit entangling gates (CNOTs). 

\subsubsection{Discrete} In this type of architecture, the probability distribution is learned on a hyper-grid. The level of discretization depends on the number of qubits. If the data has dimension $d$, and $r$ qubits are used to discretize each dimension, where $r$ can be seen as the precision bits, one needs $n=rd$ qubits to create the discrete grid with $2^r\cdots 2^r = 2^{rd}$ points.

The learned state is given in Eq.~(\ref{eq:generator_quantum_state}), with the estimated probability distribution recovered via
\begin{equation}
 P_i(\boldsymbol{\theta}) =   |\langle i | G_{\boldsymbol{\theta}} \rangle|^2 =  |c_i(\boldsymbol{\theta})|^2,
 \label{eq:discrete_dist}
\end{equation}
where $|i \rangle$ is the set of basis vectors shown in Eq.~(\ref{eq:generator_quantum_state}). This means that the probability amplitudes $c_i$ directly encode the values of the learned probability distribution. A consequence is that each measurement of the state of the generator, in the computational basis, directly samples from the desired distribution. Therefore, this type of architecture allows for very efficient sampling, only a single shot evaluation is required per sample. This is an advantage compared to the continuous architecture, where many measurement shots are required to estimate the continuous output. Here, sampling is used interchangeably with implementing a measurement in the computational basis. 

\begin{figure*}
    \centering
    \includegraphics[width=1\textwidth]{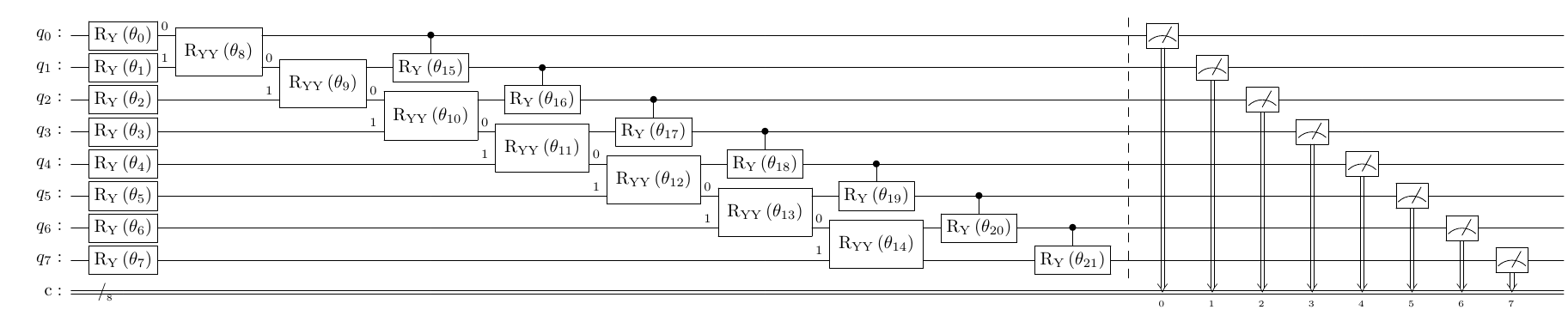}
    
    \caption{\textbf{\textbf{Discrete Standard Circuit (2D data, one block).}} This is the circuit used for both the QCBM and QGAN generators for the discrete standard case.  Each block contains 22 parametric gates. Once the circuit is applied, the state is measured in the computational basis and one shot is equivalent to one sample of the generative model, see Eq.~\eqref{eq:discrete_dist}. }
    
    \label{fig:2D_discrete_standard_circuit} 
\end{figure*}

In contrast to classical generators, a discrete generative model does not require external noise, i.\,e., all entropy comes from the quantum system. However, as we do in this work, additional external noise may be added to the output of the generator if only low numbers of qubits are available to be able to generate more than $2^n$ distinct values. We define the sampling point as the center of the grid and sample uniformly around it. This has the effect of transforming a discrete data sample into a continuous data sample. This choice has no effect on the training, and increases the diversity of the samples in the continuous space but may accentuate discretization errors.
For an example of the discrete architecture for 2D data used in this work, see Fig.~\ref{fig:2D_discrete_standard_circuit}.  Note that qubits $q_0$ to $q_3$ represent variable $x_1$ whereas qubits $q_4$ to $q_7$ represent variable $x_2$, where the output of the generative model represents the probability distribution $P(x_1, x_2)$. There is no special reason for the choice of gates, besides the attempt to limit long-range entangling interactions, which are notoriously difficult for NISQ hardware.

\begin{figure*}
    \centering
    \includegraphics[width=1\textwidth]{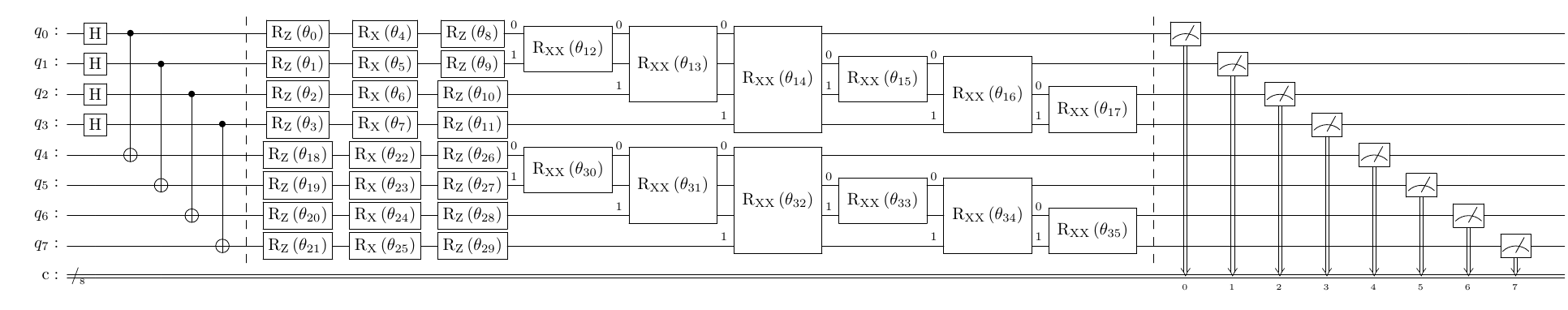}
    
    \caption{\textbf{Discrete Copula (2D data, one block).}  This is the circuit used for both the QCBM and QGAN generators for the discrete copula case. Each block contains 36 parametric gates in addition to the one-time state initialization Hadamard and C-Not gates. The purpose of the initialization stage is to create a maximally entangled state, which is then fed to the copula-structured circuit. For details, see the main text. Once the circuit is applied, the state is measured in the computational basis, and one shot is equivalent to one sample of the generative model. } 
    
    \label{fig:2D_discrete_copula_circuit} 
\end{figure*}

For this type of architecture, there exists a variant, first discussed in \cite{Zhu_2022}, referred to as \emph{copula architecture}, $U_i$,  which naturally respects the properties of PIT data, i.e., data transformed via the PIT, for details see the appendix, Sec.~\ref{appendix:quantum_copula_proof}. This means that the quantum circuit can only learn a probability distribution whose cumulative marginals are uniformly distributed, i.e., it learns the \emph{copula} of the joint probability distribution. In Fig.~\ref{fig:2D_discrete_copula_circuit}, we show the copula architecture used in this work, which is the same as proposed in \cite{Zhu_2022}. We notice that the $SU(2)$ rotations are parametrized as $R_ZR_XR_Z$, but could also be parametrized as $R_XR_ZR_X$. This choice would lead to an asatz with the same expressivity but with less training parameters, since the $R_X$ rotations can be commuted to the end of the circuit and thus get combined with the ones at the beginning of each circuit block. Here, however, we use the circuit as proposed in \cite{Zhu_2022}.  As before, qubits $q_0$ to $q_3$ represent variable $x_1$ whereas qubits $q_4$ to $q_7$ represent variable $x_2$. Note that this architecture works by first creating a maximally entangled state of the two variable registers, using Hadamard and CNOT gates, with no subsequent entangling gates between the registers. This construction ensures that, when sampling from this circuit, the marginal probabilities are uniform and thus enforces the structure of the copula. In our experiments, we pre-process the given data using the probability integral transform, as detailed in Sec.~\ref{subsec:data_transforms}, and learn its copula using the discrete copula architecture. To understand the impact the PIT has on the learning process, we also use a discrete architecture which does not respect the PIT structure. For this case, we choose an architecture used in \cite{9605352} as our standard circuit, which is shown to perform well in general. We emphasize that such choice has no relation with the data sets we use, and that other architectures which are non-PIT compliant will behave similarly.

As with the continuous case, we have simulated the discrete architecture with a state-vector simulator in which we directly simulate each shot of a measurement in the computational basis. Since one-shot per sample is needed, this choice proved to produce reasonable simulation times. 

\subsection{Benchmarking metrics}\label{sec:metrics}

In generative modeling, the task is to learn the probability distribution from which a given data set was sampled. We will concentrate on assessing the quality of the training procedure. We will not address the question of generalization of the generative model, as it is still an open problem and is beyond the scope of this work.

To understand how well the model has learned the probability distribution of the data, we can, in principle, use any distance metric between probability distributions. For example, 2-norm distance, Hellinger distance, Kolmogorov-Smirnov statistic \cite{whitnall2011exploration}, or Wasserstein distance \cite{panloss2020} are all possible choices. Here, we use the most commonly used metric capable of measuring the difference between probability distributions: the Kullback-Leibler divergence, defined in Eq.~(\ref{eq:KL_loss}). Once the training of the generative model has converged, the best model parameters $\boldsymbol{\theta}$, with respect to the Kullback-Leibler divergence between an estimate of the underlying probability distribution and the histogram of the training data, are used to generate synthetic data samples. The model's performance is measured by the value of the KL divergence.

\subsection{Data sets}\label{sec:datasets}
To understand the performance of the various combinations of model architectures, data transformations, and training methods, we learn the probability distributions of synthetic and real data sets.  We restrict ourselves to low-dimension data (2 and 3 dimensions) as simulating the training process in classical computers is very time-consuming. The data sets we use are:
\begin{itemize}
    \item Mixture of Gaussian distributions (MG):  We randomly create a 50/50 mixture of data points sampled from two Gaussian distributions with random means and covariances. 
    \item X: We create a data set where points are randomly sampled from an X-shaped distribution, as a more extreme mixed distribution. This distribution has uniform marginals.
    \item O:  We create a data set where points are uniformly distributed on the unit circle and the surface of the unit sphere.
    \item Stocks:  We take time series stock pricing from Yahoo! Finance \cite{yfinance} and compute the daily returns, which is the percentage change of the close price between consecutive days.
\end{itemize}

\begin{figure*}[t]  % 2d  
    \centering
     \includegraphics[width=1\textwidth]{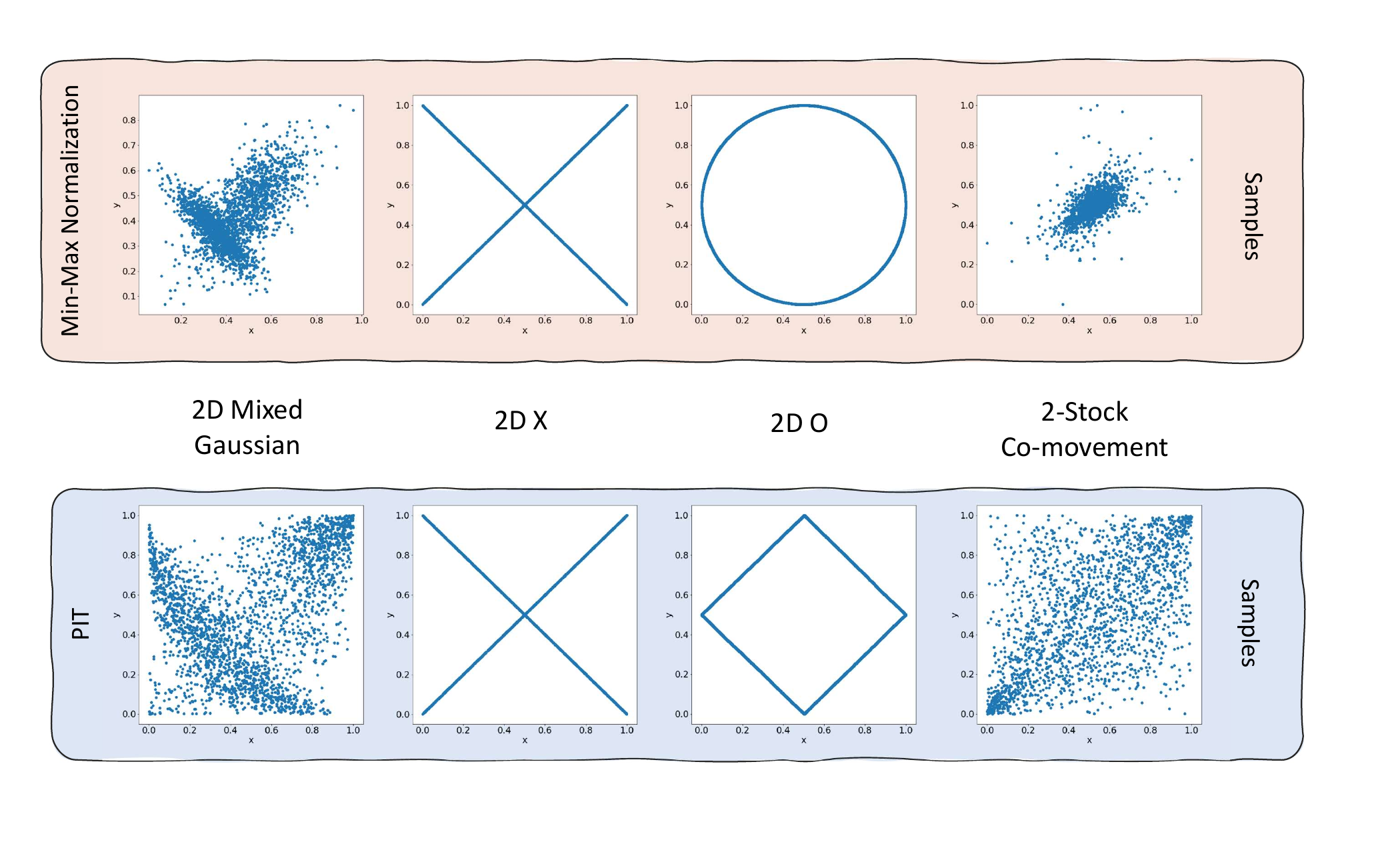}
    \caption{\textbf{Normalized 2D data sets.} Top (orange) min-maxed normalized samples, bottom (blue) are the PIT data sets.} \label{fig:data_visulaization_2d}
\end{figure*}

\begin{figure*}[t] % 3d
    \centering  
    \includegraphics[width=1\textwidth]{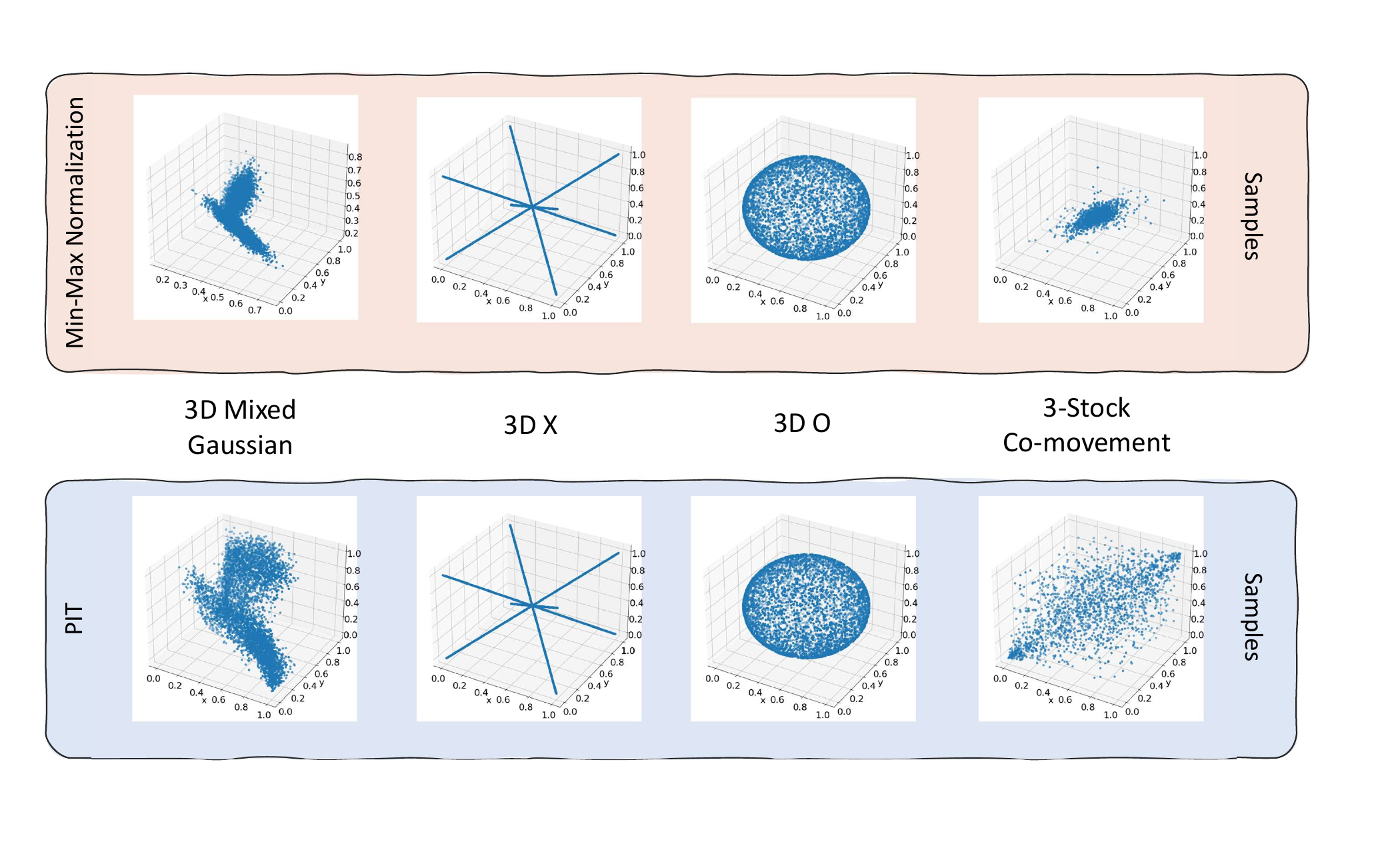}
    \caption{\textbf{Normalized 3D data sets.} Top (orange) min-maxed normalized samples, bottom (blue) are the PIT data sets.}\label{fig:data_visulaization_3d}
\end{figure*}

These data sets were selected to have a variety of different properties, for instance, uniform and non-uniform marginals, or specific means and co-variances. These synthetic data sets have the advantage that they can be processed by the candidate quantum architectures without further compression or auto-encoding. Except for the stocks data, all the 2D distributions had $50{,}000$ data points and the 3D distributions had $100{,}000$ data points to try to compensate for the curse of dimensionality. The stocks data for both 2D and 3D had $1{,}877$ data points. We show samples and histogram representations of the four data sets used in this work in Figs.~\ref{fig:data_visulaization_2d} and \ref{fig:data_visulaization_3d}. Note that the mixed Gaussian and stocks data are shifted and distorted when the PIT is applied, as expected. However, the X and the O 3D only get shifted to the first quadrant and octant, respectively. This is because their original distributions already have uniform marginals. In contrast, the O 2D case is interesting, as it gets both shifted and distorted when the PIT is applied. We include an explicit calculation of the O 2D case for the interested reader in Appendix~\ref{appendix:pit_calculation}. 

\section{Numerical Results}\label{sec:results}

In this section, we present the results of our experimentation using the described architectures and data transformations. We train and evaluate the performance of our quantum generative models using the KL divergence, described in Sec.~\ref{sec:metrics}.
We aim to understand how the learning and training behavior compares when the complexity of the quantum and classical networks are varied to find indications of possible quantum advantage. We vary the number of training parameters $\boldsymbol{\theta}$ for each model as we study their scaling behavior. 
The number of required parameters to achieve a certain accuracy ultimately affects performance and training run time. We choose the number of model parameters as a fair indicator of model size, as the concept of circuit depth and width is not necessarily comparable between classical and quantum models.  

\begin{figure*}[t]
 \begin{subfigure}{\textwidth}
    \centering
        \vspace{0.5cm}
        \caption{\textbf{Classical Models}}
        \includegraphics[width=\textwidth]{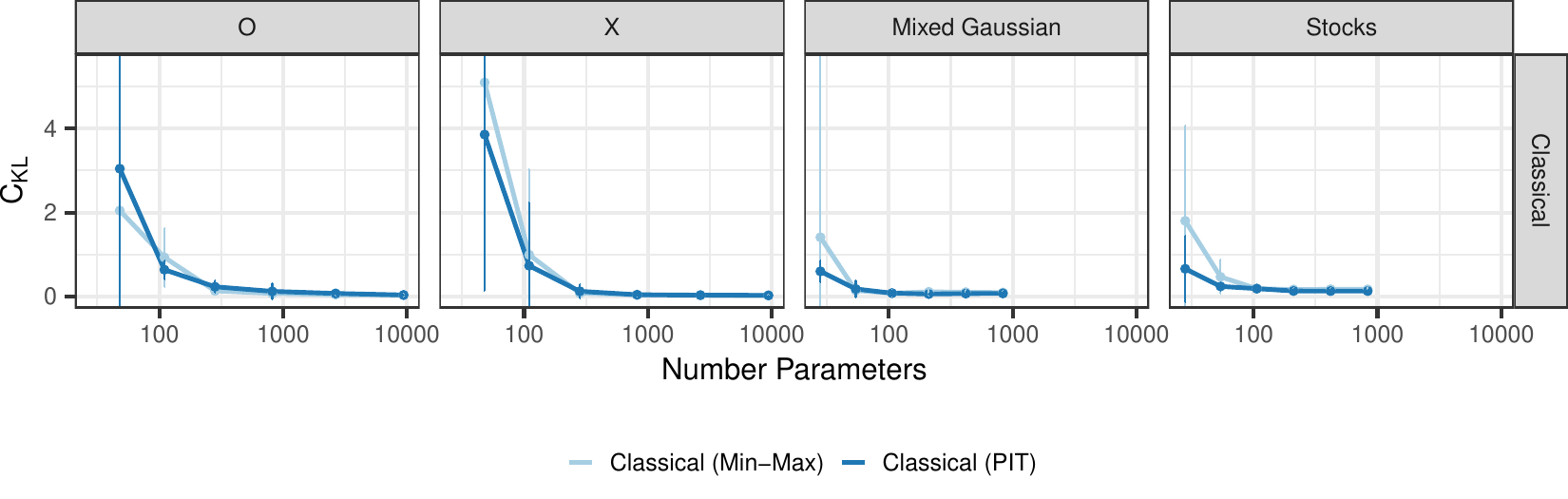}
        
        \label{fig:2D_results_classic}
    \end{subfigure}
    \begin{subfigure}{\textwidth}
    \centering
        \caption{\textbf{Quantum Models} }
        \includegraphics[width=\textwidth]{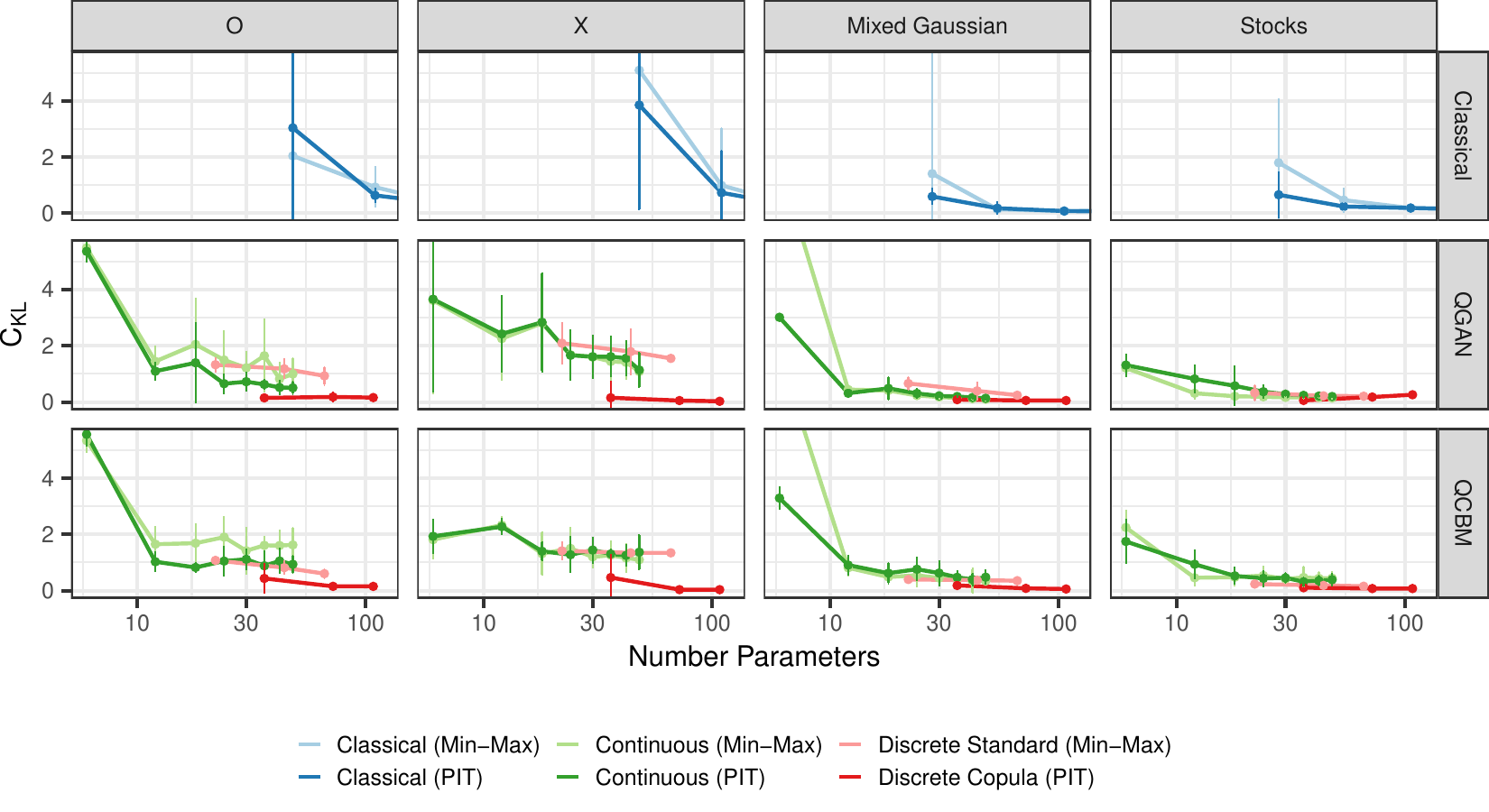}
         
        \label{fig:2D_results_quantum} 
    \end{subfigure}

 \caption{\textbf{KL Divergence for Different Parameters, Model Architectures, Data Normalization and Training for 2D data}. Each line shows the average KL divergence over 5 experimental runs of each model for each parameter count, with the error bars given by the standard deviation over 5 runs. a) shows only the classical model results and b) compares classical and quantum results. The x-axis on both figures are shown in logarithmic scale but cover different ranges. Note that only for the X data set, min-max normalization and PIT coincide, and therefore for the continuous and classical architectures their KL divergences overlap.}
    \label{fig:fig_KL_results}
\end{figure*}

\begin{figure*}[t]
    \begin{subfigure}{\textwidth}
    \centering
        \vspace{0.5cm}
        \caption{\textbf{Classical Models}}
        \includegraphics[width=\textwidth]{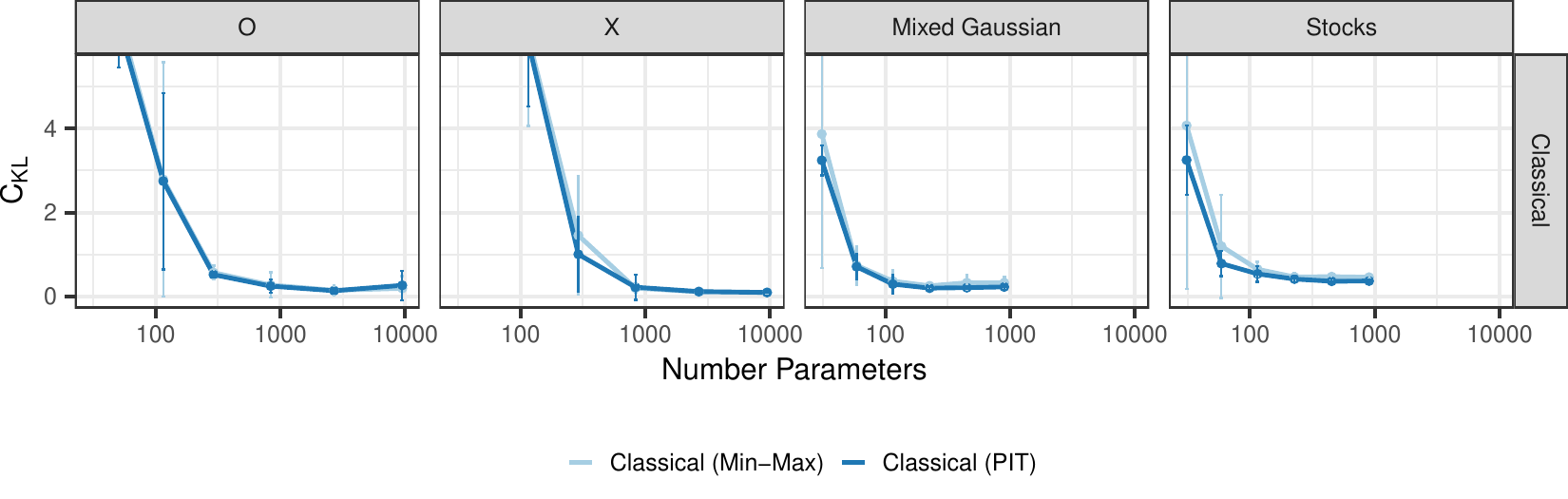}
        \label{fig:3D_results_classic}
    \end{subfigure}
     \begin{subfigure}{\textwidth}
    \centering
        \caption{\textbf{Quantum Models} }
       \includegraphics[width=\textwidth]{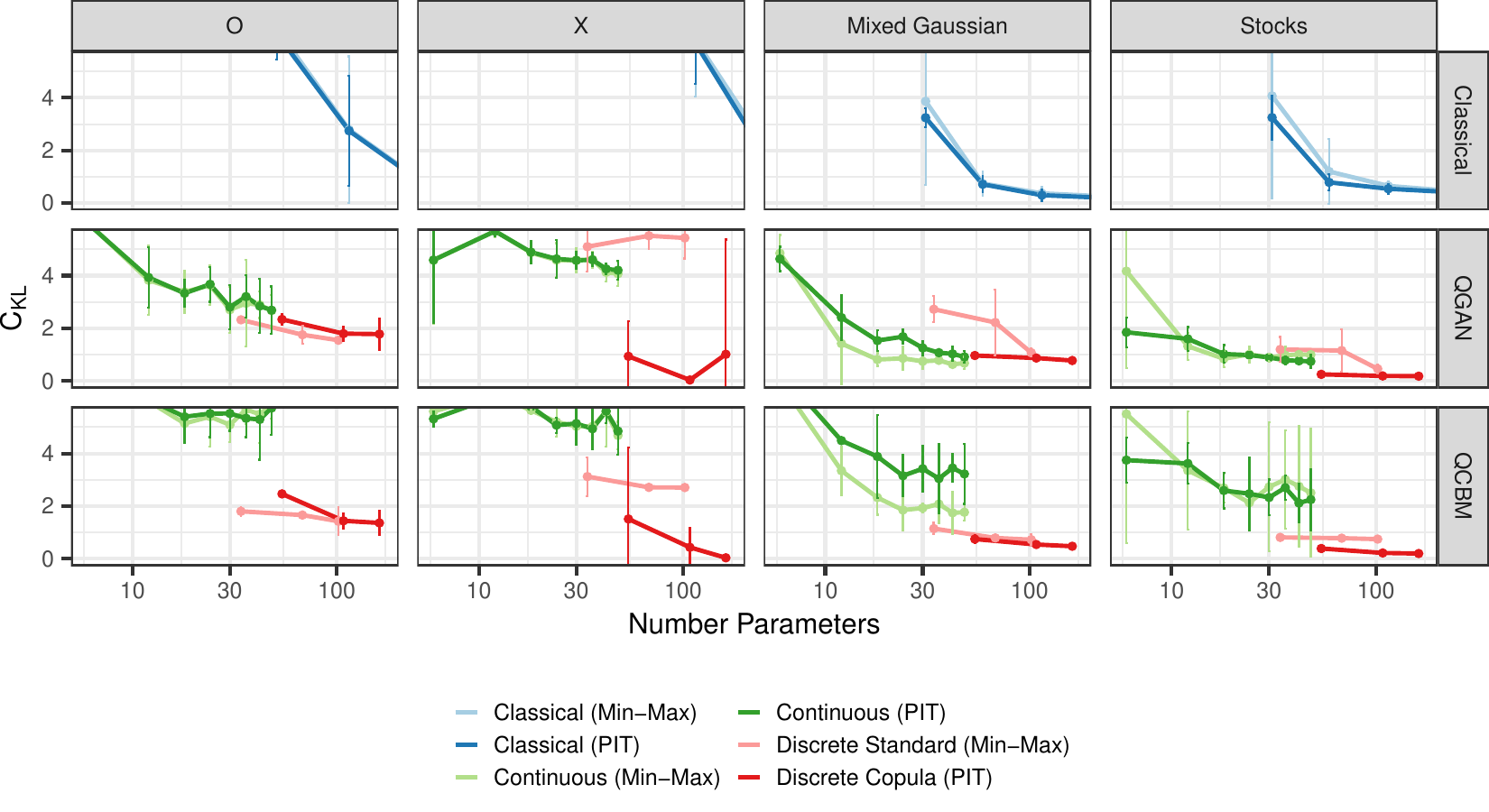} 
        \label{fig:3D_results_quantum} 
    \end{subfigure}
 \caption{\textbf{KL Divergence for Different Parameters, Model Architectures, Data Normalization and Training for 3D data}. Each line shows the average KL divergence over 5 experimental runs of each model for each parameter count, with the error bars given by the standard deviation over 5 runs. a) shows only the classical model results and b) compares classical and quantum results. The x axis on both figures are shown in logarithmic scale but cover different ranges. Note that only for the X  and O data sets, min-max normalization and PIT coincide, and therefore for the continuous and classical architectures their KL divergences overlap.}
    \label{fig:fig_KL_results_3D}
\end{figure*}

For quantum models, the number of parameters was varied by changing the depth (number of blocks) of the parametric circuits (see Figs.~\ref{fig:2D_continuous_circuit},  \ref{fig:2D_discrete_standard_circuit}, and \ref{fig:2D_discrete_copula_circuit}). For classical models, their size was changed by varying the number and width of the hidden layers of a neural network. Note that due to the differences in model architectures, the number of parameters across our experiments does not start at the same point or increase in the same manner. We repeat each experiment five times to produce statistics of the training performance and estimate the standard deviation of the KL divergence for classical GANs, QCBMs, and QGANs. We used classical state vector simulators for the quantum models and a GPU-accelerated computer for the classical models. 
%Figure~\ref{fig:2D_results_quantum} shows the KL divergence for the quantum models and figure~\ref{fig:2D_results_classic}. 
Tables \ref{tab:min_kl_2D} and \ref{tab:min_kl_3D} show the lowest KL value over all experimental runs for each model and data set and the number of parameters used for that result. Figures \ref{fig:fig_KL_results} and \ref{fig:fig_KL_results_3D} show the average KL divergence found over five runs for every model type and data set.

A trend seen over all data sets and model varieties was that increasing the number of parameters decreased the KL divergence.  These experiments do not demonstrate that simulated quantum circuits go beyond the current capabilities of classical GANs. However, they do show that there are instances where quantum circuits require far fewer parameters to achieve a similar KL divergence. For example, for the 3D X data set, a classical GAN with a min-max normalization with 9475 parameters achieves a minimum KL of 0.089, whereas a QGAN with a discrete copula architecture, PIT normalization, and 54 parameters achieves a KL of 0.029. We see that for the O and X data sets for both 2D and 3D, all quantum models performed better than the classical models with similar numbers of parameters. The quantum models perform comparably close to the classical ones at similar parameter counts for the MG and stocks data, with the discrete copula models consistently outperforming the classical ones. 

%Table~
\begin{table*}

\caption{\textbf{Minimal KL Divergence and Number of Parameters for 2D Data.} Figures in bold are the minimum KL values found across all training methods for each data set. \label{tab:min_kl_2D}}
\centering
\resizebox{\textwidth}{!}{
\begin{tabular}[t]{|lll|cc|cc|cc|cc|}
% \begin{tabular}[t]{lllrrrrrrrr}
\hline
\hline
% \toprule
% \multicolumn{3}{c}{Configuration } & \multicolumn{8}{c}{Dataset, KL Divergence and Number of Parameters} \\
\multicolumn{3}{|c}{Configuration } & \multicolumn{8}{|c|}{Dataset, KL Divergence and Number of Parameters} \\
\hline
% \cmidrule(l{3pt}r{3pt}){1-3} \cmidrule(l{3pt}r{3pt}){4-11}
Circuit & Training & Normalization & MG KL & MG Params & O KL & O Params & Stocks KL & Stocks Params & X KL & X Params\\
\hline
% \midrule
\cellcolor{gray!6}{Classical} & \cellcolor{gray!6}{Classical} & \cellcolor{gray!6}{Min-Max} & \cellcolor{gray!6}{\textbf{0.040}} & \cellcolor{gray!6}{106} & \cellcolor{gray!6}{\textbf{0.022}} & \cellcolor{gray!6}{9410} & \cellcolor{gray!6}{0.145} & \cellcolor{gray!6}{210} & \cellcolor{gray!6}{0.020} & \cellcolor{gray!6}{9410}\\
Classical & Classical & PIT & 0.054 & 834 & 0.027 & 9410 & 0.101 & 418 & 0.023 & 2658\\
\cellcolor{gray!6}{Continuous} & \cellcolor{gray!6}{QGAN} & \cellcolor{gray!6}{Min-Max} & \cellcolor{gray!6}{0.083} & \cellcolor{gray!6}{48} & \cellcolor{gray!6}{0.339} & \cellcolor{gray!6}{42} & \cellcolor{gray!6}{0.117} & \cellcolor{gray!6}{48} & \cellcolor{gray!6}{0.889} & \cellcolor{gray!6}{36}\\
Continuous & QGAN & PIT & 0.109 & 48 & 0.328 & 48 & 0.179 & 48 & 0.941 & 48\\
\cellcolor{gray!6}{Continuous} & \cellcolor{gray!6}{QCBM} & \cellcolor{gray!6}{Min-Max} & \cellcolor{gray!6}{0.198} & \cellcolor{gray!6}{42} & \cellcolor{gray!6}{1.031} & \cellcolor{gray!6}{30} & \cellcolor{gray!6}{0.271} & \cellcolor{gray!6}{18} & \cellcolor{gray!6}{0.897} & \cellcolor{gray!6}{42}\\
Continuous & QCBM & PIT & 0.294 & 48 & 0.499 & 36 & 0.235 & 36 & 0.800 & 24\\
\cellcolor{gray!6}{Discrete Copula} & \cellcolor{gray!6}{QGAN} & \cellcolor{gray!6}{PIT} & \cellcolor{gray!6}{0.047} & \cellcolor{gray!6}{72} & \cellcolor{gray!6}{0.142} & \cellcolor{gray!6}{108} & \cellcolor{gray!6}{\textbf{0.048}} & \cellcolor{gray!6}{36} & \cellcolor{gray!6}{\textbf{0.019}} & \cellcolor{gray!6}{72}\\
Discrete Copula & QCBM & PIT & 0.053 & 108 & 0.147 & 72 & 0.056 & 72 & 0.032 & 72\\
\cellcolor{gray!6}{Discrete Standard} & \cellcolor{gray!6}{QGAN} & \cellcolor{gray!6}{Min-Max} & \cellcolor{gray!6}{0.212} & \cellcolor{gray!6}{66} & \cellcolor{gray!6}{0.731} & \cellcolor{gray!6}{66} & \cellcolor{gray!6}{0.192} & \cellcolor{gray!6}{66} & \cellcolor{gray!6}{1.399} & \cellcolor{gray!6}{44}\\
Discrete Standard & QCBM & Min-Max & 0.325 & 66 & 0.513 & 66 & 0.131 & 66 & 1.337 & 44\\
% \bottomrule
\hline
\hline
\end{tabular}}
\end{table*}

\begin{table*}

\caption{\textbf{Minimal KL Divergence and Number of Parameters for 3D Data.} Figures in bold are the minimum KL values found across all training methods for each data set. \label{tab:min_kl_3D}}
\centering
\resizebox{\textwidth}{!}{
\begin{tabular}[t]{|lll|cc|cc|cc|cc|}
 \hline
 \hline
\multicolumn{3}{|c}{Configuration } & \multicolumn{8}{|c|}{Dataset, KL Divergence and Number of Parameters} \\
 \hline
% \cmidrule(l{3pt}r{3pt}){1-3} \cmidrule(l{3pt}r{3pt}){4-11}
Circuit & Training & Normalization & MG KL & MG Params & O KL & O Params & Stocks KL & Stocks Params & X KL & X Params\\
\hline
\midrule
\cellcolor{gray!6}{Classical} & \cellcolor{gray!6}{Classical} & \cellcolor{gray!6}{Min-Max} & \cellcolor{gray!6}{0.193} & \cellcolor{gray!6}{227} & \cellcolor{gray!6}{0.075} & \cellcolor{gray!6}{9475} & \cellcolor{gray!6}{0.417} & \cellcolor{gray!6}{227} & \cellcolor{gray!6}{0.089} & \cellcolor{gray!6}{9475}\\
Classical & Classical & PIT & \textbf{0.167} & 227 & \textbf{0.069} & 9475 & 0.339 & 451 & 0.091 & 9475\\
\cellcolor{gray!6}{Continuous} & \cellcolor{gray!6}{QGAN} & \cellcolor{gray!6}{Min-Max} & \cellcolor{gray!6}{0.543} & \cellcolor{gray!6}{48} & \cellcolor{gray!6}{1.518} & \cellcolor{gray!6}{36} & \cellcolor{gray!6}{0.660} & \cellcolor{gray!6}{18} & \cellcolor{gray!6}{2.949} & \cellcolor{gray!6}{6}\\
Continuous & QGAN & PIT & 0.808 & 48 & 2.045 & 42 & 0.575 & 48 & 2.978 & 6\\
\cellcolor{gray!6}{Continuous} & \cellcolor{gray!6}{QCBM} & \cellcolor{gray!6}{Min-Max} & \cellcolor{gray!6}{1.388} & \cellcolor{gray!6}{42} & \cellcolor{gray!6}{4.557} & \cellcolor{gray!6}{24} & \cellcolor{gray!6}{0.876} & \cellcolor{gray!6}{42} & \cellcolor{gray!6}{4.283} & \cellcolor{gray!6}{48}\\
Continuous & QCBM & PIT & 2.324 & 36 & 4.160 & 42 & 1.223 & 48 & 4.115 & 48\\
\cellcolor{gray!6}{Discrete Copula} & \cellcolor{gray!6}{QGAN} & \cellcolor{gray!6}{PIT} & \cellcolor{gray!6}{0.726} & \cellcolor{gray!6}{162} & \cellcolor{gray!6}{1.315} & \cellcolor{gray!6}{162} & \cellcolor{gray!6}{\textbf{0.168}} & \cellcolor{gray!6}{162} & \cellcolor{gray!6}{\textbf{0.029}} & \cellcolor{gray!6}{54}\\
Discrete Copula & QCBM & PIT & 0.460 & 162 & 1.006 & 162 & 0.188 & 162 & \textbf{0.029} & 162\\
\cellcolor{gray!6}{Discrete Standard} & \cellcolor{gray!6}{QGAN} & \cellcolor{gray!6}{Min-Max} & \cellcolor{gray!6}{0.990} & \cellcolor{gray!6}{102} & \cellcolor{gray!6}{1.514} & \cellcolor{gray!6}{102} & \cellcolor{gray!6}{0.354} & \cellcolor{gray!6}{102} & \cellcolor{gray!6}{4.636} & \cellcolor{gray!6}{34}\\
Discrete Standard & QCBM & Min-Max & 0.645 & 102 & 1.112 & 102 & 0.686 & 102 & 2.692 & 68\\
% \bottomrule
 \hline
 \hline
\end{tabular} }
\end{table*}

\subsection{Classical GAN training results}

We train a classical GAN using the data sets described in Sec.~\ref{sec:datasets} for empirical comparison with the results of our quantum generative models. 
We used a fully connected neural network architecture and varied its details to find the best classical GANs to learn the data distributions.
In our experiments, we varied the number of input parameters, depth (number of layers), and width (size of the hidden layers) of the neural network. 

We found that for the mixed-Gaussian and stocks data sets 2 hidden layers were sufficient to explain the data, while for the X and O data sets 4 hidden layers were needed. An input noise dimension of 8 was found to work well and is used in all models discussed here. Reducing the input noise dimension further is possible, but at the expense of a higher KL divergence. In Figure~\ref{fig:fig_KL_results}, the number of training parameters is varied by changing the size of the hidden layers from 2 to 64.

The performance of the classical models for all 2D data sets is comparable to those of the good quantum models. For the 3D case, this is true except for the O data set, where the quantum models struggle. For all data sets, the necessary number of parameters to achieve a similar KL divergence is generally higher for the classical models than for their quantum counterparts. While this difference is moderate for the mixed Gaussian and stocks data sets, for the O and X data sets the classical model required two orders of magnitude more parameters to achieve the same KL divergence as the best quantum models. This suggests that the quantum circuits need fewer parameters to learn more complex data. 

\subsection{QCBM training results}
We start our QML experimentation by training a QCBM using the data sets defined in Sec.~\ref{sec:datasets}. For the discrete architectures, we discretize each dimension using 4 qubits. This means that we need 8 (12) qubits for 2-dimensional (3-dimensional) data. For the continuous architecture, we use 1 qubit per dimension. For all architectures, we minimized Eq.~\eqref{eq:kl_optimization} via a state-of-the-art gradient-free method, CMA-ES \cite{hansen2019pycma}. As the number of parameters of the model increases, training becomes more challenging, i.\,e., it is harder to find a suitable set of model parameters that can recreate the target distribution. In our tests, we create identical blocks of quantum gates and add them one after another, which linearly scales the number of parameters and is common practice. This allows us to initialize the subsequent block with the fitted parameters of the previously trained blocks, while the new block's parameters are initialized to zero, or equivalently the new block's unitary is initialized to the identity operator. In training, each subsequent block is thereby treated as a correction unitary, which brings us closer to the limit of what the quantum circuit is capable of learning. We find that this approach mitigates convergence problems compared to trying to fit all free parameters simultaneously every time.

Overall, for the majority of the data sets (6 of 8) the variations in performance due to architecture and normalization had a greater effect than the choice of QGAN or QCBM model. For the 2D data sets, the relative performance of QGAN and QCBM models is similar, although the best KL divergence from a quantum model was from a QGAN for all four data sets. In the 3D case, the QGAN model outperformed the QCBM models for the mixed Gaussian and stocks data sets and performance was comparable for the X and O data sets.

The discrete-type architecture gave better results than the continuous circuit for the 2D data sets, except for the X data set, where only the discrete copula circuit performed well, not the standard circuit. For the 3D data sets, the discrete architecture outperformed the continuous for all four data sets. 

For the 2D and 3D data sets, the PIT normalization outperformed the min-max normalization for the discrete architecture. In particular, there is a clear advantage in using a discrete-type architecture with the copula circuit for 2D the X and O data sets and for X 3D.
We believe this is because the PIT simplifies the correlations of the data sets, and the fact that the quantum architecture enforces a copula-type data structure, as described in the appendix, Sec.~\ref{appendix:quantum_copula_proof}. Specifically, we see that the standard discrete architecture struggles with the X data set for both 2D and 3D. This could be because this data set has naturally uniform marginals, which the standard circuit is not equipped to handle natively, and thus must learn. The O 2D data set also seems to present some complications for the discrete standard architecture and the continuous one. Conversely, the performance difference between PIT and min-max normalization for the MG and Stocks data is smaller. We suggest that this is because these data sets both share Gaussian properties, as shown in Fig.~\ref{fig:data_visulaization_2d}, and hence and simple correlations, so simplifying them further with a PIT normalization only makes a marginal difference.

For the continuous architecture, min-max and PIT performed similarly, except for the 2D O and X data sets where PIT was better, and for the 3D mixed Gaussian where the min-max was better. We suggest that may be true because the same quantum circuit structure is used for both types of normalization. 
  
The parameter scaling for the PIT data with the discrete copula circuit is favorable. For 2D data, even for a low number of parameters (only 1 circuit block), the results are already good across all data sets.  For both 2D and 3D data, we do not see a significant improvement in the minimum value of KL found as the additional blocks are added. This indicates that learning copula data is more accessible than learning the full joint probability distribution, as done for the min-max normalization case.

\subsection{QGAN training results}
We continue our QML experimentation by training a QGAN using the same data sets as described in Sec.~\ref{sec:datasets}. As in the case of the QCBM training, the simulations are run in the same discretized space (8 and 12 qubit circuits) for the discrete architectures and 1 qubit per data dimension (2 and 3 qubits) for the continuous ansatz. The setup follows the one introduced in \cite{NIPS2014_5ca3e9b1} but with a quantum generator and with the classical discriminator differing between the continuous and discrete cases, see Fig. \ref{QGAN_training_combined}.  
For the continuous, the discriminator is a neural network with 2 input features for 2D data and 3 for 3D,  and two hidden layers with 32 nodes each, after which ReLU activations are applied. The last layer contains a single output node with a sigmoid activation function.  Note that in the continuous case, all experiments use the same architecture for both normalization methods. The discrete experiments also use a neural network discriminator, but in contrast to the continuous case, it takes as input bit strings of dimension 8 (12) for 2D (3D) data and a single hidden layer with 16 (24) nodes followed by leaky ReLu activation functions. As in the continuous case, the last layer produces a single output followed by a sigmoid function to ensure normalization.
 For training of all QGANs we do not use lower-depth trained circuits to initialize the training of high-depth circuits (i.e., we cold-started all QGANs), as achieving convergence of the individual QGANs was not an issue.  All QGANs were trained by gradient descent methods. Our implementation of the continuous QGAN architecture used Optax, which is a package for convenient computation of gradient-descent optimization, built on the automatic differentiation engine provided by JAX \cite{deepmind2020jax}. For the discrete architectures, we used a gradient descent implementation based on the parameter-shift rule, see for example \cite{Mitarai_2018}.

Overall, QGAN performs slightly better than the QCBM for both 2D and 3D data, although the differences are smaller than the variations due to circuit architecture and normalization.
In the 2D case,  the discrete copula  architecture with a PIT normalization was the most performant, and the standard discrete architecture and both continuous models were worse. The difference in performance was mostly for the O and X data sets.  
In the 3D case, the situation is different. The continuous circuits performed similarly or better than the discrete for the O, MG and Stocks data sets. For the X data set the continuous  models perform poorly compared to the discrete PIT architecture but better than the discrete min-max circuit.  

Interestingly, the continuous architecture does not benefit from the PIT at the same level as the discrete architecture.  This could be due to several factors: 1.\,The specific choice of hyper-parameters could be favoring the min-max normalization. 2.\,The quantum circuit ansatz for the continuous GAN is the same for PIT and min-max normalization, which is not true for the discrete QGAN (as is the case for the QCBM). 3.\,The circuits might not have been deep enough. 4.\,It still needs to learn the structure of the marginal distributions, which is not enforced by the circuit architecture.

\section{Conclusions and outlook}\label{sec:discussion}

\subsection{Quantum generative modeling}
As quantum states are natural generators of probability distributions, quantum computing seems poised to provide advantages in the field of generative modeling. In this work, we observed that with relatively simple circuit architectures, it is possible to learn a variety of different data sets.
For most data sets considered, the quantum models required fewer parameters to achieve the same or better performance than similarly sized classical counterparts. These are encouraging results. We have also found that performance can vary greatly depending on circuit architecture and normalization choices, indicating that these are important factors to consider when developing new generative models. Although we did not find a one-size-fits-all model, using the  discrete copula architecture, trained with either a QCBM or QGAN, consistently performed well. 

In our tests, we found that the stability of training a QGAN is highly dependent on the model architecture, see Appendix \ref{appendix:training_curves}. When using PIT data, we observed fewer mode collapses or similar instabilities, that are common to GANs, when compared with min-max normalized data for the discrete copula architecture. This is an indication of the simpler nature of learning the copula of a probability distribution and the enforced copula structure by the quantum circuit. 
For the continuous architecture, however, we do not see any difference in behavior between min-max normalized data and PIT data, which may be explained by the fact that the quantum circuit structure is generic and does not enforce the copula structure.
More investigations are needed to make a definite conclusion. We believe that further enhancements are possible by modifying the training process of the QGANs, for example using a quantum Wasserstein GAN~\cite{arjovsky2017}, which we will explore in future work. 

From the application perspective, generative models can generally be used as synthetic data generators for cases in which real data is scarce for training AI systems.
However, as pointed in \cite{Zoufal2019quantum, Woerner2019}, quantum generative models can also be used as approximated data loaders for quantum information processing tasks. For example, in finance, they can be used to compute the pricing or analyze the risk of an asset. Exploring these lines of applications is important to ensure the future industrial and commercial use of quantum technologies. 
We simulated all our circuits on classical computers, to extend our understanding of the limitations of quantum technologies, future research must also explore the effects of decoherence in generative modeling. With regards to finite sampling effects; for the discrete architectures, the output is itself a single measurement outcome. In fact, in the simulations, we are sampling from the probability amplitudes of the state vectors to generate the output of the model. In this case, it is clear that finite statistical effects are present. For the continuous case, the continuous outputs remain deterministic \textit{for a fixed input random noise vector,}  $\mathbf{z}$, i.e., the exact expectation value is computed, but are probabilistic due to this random variable vector. In this case, even if the exact expectation value is computed as the circuit output, there are finite statistical effects as a finite number of noise vectors are used during training. In this case, there is a distinction between quantum shot-noise effects and the finite statistical effects. For completeness, please see Appendix C, Table III where we comment on the training behavior if finite samples are used when training the continuous architectures, regarding runtime of the algorithm.
For a low qubit count, i.e. a coarse grid, covering the probability space is possible with reasonable resources and is in fact a requirement for the correctness of the QCBM training procedure. Further work could discuss the effects of sample size for higher-width quantum circuits, especially using the QGAN training paradigm.

\subsection{Generalization and scaling}
In machine learning, it is crucial to determine the performance of a trained model. There are two aspects to assess: 1.\,the quality of the training, i.\,e., how well the model reproduces the training data, and 2.\,the quality of the generalization, i.\,e., how well the model can generate data that was not present in the training set. The study of the generalization of machine learning models is a very active research area both in the quantum and classical cases, e.\,g., \cite{DBLP:journals/corr/abs-2201-08770, broji2018, broji2021, ermon2018, gili2022generalization}. In this work, we have focused on the first aspect, using the KL divergence to measure the quality of the training, i.\,e., we choose the model which produces samples that are closest to the probability distribution of the training data. In the context of generative modeling, this does not necessarily lead to over-fitting, as the intention is to fit the input distribution as well as possible without memorizing the original data. However, for real-world applications, e.\,g., in synthetic data generation, particularly with higher dimensional data, assessing the quality of generalization is important and should be addressed in further work. Moreover, we have restricted ourselves to dealing with 2D and 3D data sets. In future work, we will explore the interplay of learning processes and generalization with higher dimensional data. In particular, to see if the empirical advantage found in this work in the number of parameters required can be carried over to higher dimensions.

\subsection{Model architectures}

In this work, we simulated the training of many generative models to better understand the impact of model type and architecture on performance. We found that learning a data set that has been transformed via PIT paired with the discrete copula ansatz gives, for most data sets, the best results for both QCBM and QGAN. This is due to the simpler correlations the PIT data presents, compared to the original probability distributions, and the structure of the copula quantum circuit. 

We see a clear distinction between the continuous and discrete architectures. The continuous ansatz is capable of producing continuous valued samples and has the advantage of requiring fewer qubits, making it more accessible in the NISQ era. The cost of these advantages is that the output of this type of model is computed via estimation of expectation values. When run on quantum hardware, the training run-time is expected to be long, as many shots of the quantum circuit are needed to evaluate even a single data sample.

In our experiments, we found that in the majority of cases, the continuous architecture did not produce the best results. However, we only explored a small range of all the possible data encodings, cost functions, hyper-parameters and training techniques, so better results may be possible.

The discrete architecture is restricted to producing samples in a discrete grid. In our tests, we discretized each dimension in 16 segments, which is too low a resolution for most real-world applications. This is the limiting factor for NISQ era hardware, as the discretization level depends on the number of available qubits. To generate more than $2^n$ distinct samples on a NISQ system, we artificially add uniformly distributed noise to each discrete sample. The main advantage of the discrete architecture is that it is more measurement-, and thus, will be more runtime-efficient on a real quantum system for training and sampling (see Table \ref{tab:resource_comparison} in the appendix). As each measurement can be directly interpreted as a sample, it is both runtime and resource-efficient.

In contrast to a hardware implementation, the simulations of the continuous architectures are much faster to train than the discrete architectures for two reasons: the output is computed directly from state vector simulation, so only one evaluation is needed per sample and, as mentioned above, only one qubit per dimension is simulated. Conversely, the discrete circuits use more qubits and the simulated outcomes are computed via direct sampling of the state vector, which is, in general, more costly to simulate. The advantage of the continuous architecture does not apply to real quantum hardware where multiple measurements are required.
 
The choice of the precise circuit architecture for generative models we made in this paper follows similar practices in the literature but is still quite arbitrary, mainly dictated by the connectivity and gate availability of particular hardware systems. There is currently a great interest in tensor-network-inspired architectures, which not only have the advantage of potentially needing less qubit connectivity but also have enough expressive power to be useful in practice. Additionally, they are amenable to circuit cutting techniques, see for example \cite{guala_tensornetworks_qml, xanadu_circuitcutting}, which can extend the size of the circuits which can be run in NISQ era hardware via a classical computational overhead.  In the future, we will explore whether tensor networks can scale our architectures and models, and can learn suitable representations of real-world data. 

\subsection{Near term industrial applicability} 

The number of qubits and the circuit depth required for our quantum models are comparable to the capabilities of existing quantum computers, at maximum using $12$ qubits and $174$ gates. For example \cite{huang2021experimental} presents an implementation of a QGAN on quantum hardware using 9 trainable parameters for the quantum generator and 12 for their quantum discriminator and a maximum of 6 qubits. Although this parameters are lower than the simulations presented in this work, we believe they are close enough to claim are models are within the NISQ regime.  Although the data sets considered here are low dimensional, industrially relevant scales for certain applications are not far off. 
For example, instances of sampling from probability distributions can occur in healthcare Real World Evidence cases. In these scenarios, a set of qubits can be used to represent values of integer or categorical variables, which are then aggregated to encode a data point of the probability distribution. This process is particularly applicable to real-world healthcare datasets, as illustrated in \cite{Hossain_2024}, where patient characteristics (such as gender) and variables like symptoms (e.g., chest pain, diagnosis, etc.) are categorized, some of which include integer variables (e.g., age, blood pressure) in a limited range that can be encoded using few qubits. Subsequently, the individual values of a patient can be described by a vector with approximately 50 qubits\footnote{The qubit budget would allow for 5 categorical and 8 numeric variables  assuming  two qubits per  categorical variable and  five qubits per numeric variable.}.

In future work, we will explore how these simulator-trained models perform on commercially available quantum hardware to see how much the output of such models changes in a realistic setting. The difference will indicate the amount of hardware noise on real hardware and can be used as a snapshot of the current state of quantum hardware. 

For large-scale industrial applications, though, quantum hardware still has a long way to go. Using the estimates shown in Table \ref{tab:resource_comparison} in the appendix, we can calculate the resources needed for real-life applications. Typical examples include: 1) Modeling of financial securities, whose data sets are in the range of $d=100$ dimensions, will require $100$ qubits and $400$ gates for the continuous architecture and $1600$ qubits and over $52000$ gates for the discrete copula architecture, with a modest discretization precision ($r=16$). 2) Generation of virtual patients for early-phase clinical trials in health care applications, whose data sets are also in
the order d = 100 dimensions, see for example \cite{Sinisi_Alimguzhin_Mancini_Tronci_Leeners_2020}, 
and thus similar resources than the previous example, and 3) Generation of synthetic images \cite{in-painting2022} which can be used for privacy-protection and anonymization applications. In such cases, the dimensionality of the data is of the order of $d=10^5$, which will require order $10^5$ qubits and $10^6$ gates for the continuous architecture, and $10^6$ qubits and order $10^7$ gates for the discrete architectures.

\section{Acknowledgments}
We thank our respective companies for their support. We thank Tom Endres from Accenture for supporting the operational setup of the QUTAC working group `Production \& Logisitics'. We thank Manuel Proissl for valuable discussions on industry perspectives and support of building the underlying code and effective collaboration. We also thank Thomas Strohm,  Andrea Skolik, and Jernej Rudi Finzgar for useful comments and discussions. 

    \section{Contributions}
All authors contributed to conceptualization, validation of the results and reviewing and editing the draft. 
CAR and OM contributed to the methodology and experiment design. CAR, OM, JD, AV, and FK wrote the software. CAR contributed to the formal analysis. 
 AL, TE, and JK curated and visualized the data. CAR and CJ wrote the initial draft. CAR handled supervision. JK organized the project administration.

\bibliography{literature}

\section{Appendix}

\subsection{Uniform marginals of the copula architecture}\label{appendix:quantum_copula_proof}
In this section, we will show that the copula architecture is  capable of producing samples whose marginal distributions are uniformly distributed. We will analyze explicitly the 2-dimensional case. Higher dimensions are treated in the same way.

The copula architecture starts by preparing a maximally entangled state of of registers $A$ and $B$
\begin{equation}
    |\psi_{in}\rangle = \frac{1}{2^{n/4}}\sum_{i=0}^{2^{n/2}-1}|i\rangle_A\otimes|i\rangle_B
\end{equation}
where $|i\rangle$ is labeled the representation binary of $i$. We can write the density matrix corresponding to this state
\begin{equation}
\begin{split}
    \rho_{in} &= |\psi_{in}\rangle \langle\psi_{in}|\\
    &= \frac{1}{2^{n/2}}\sum_{i,j=0}^{2^{n/2}-1} |i\rangle_A \langle j|_A \otimes |i\rangle_B \langle j|_B.
\end{split}
\end{equation}
After the variational circuit is applied, the state of the system before the measurement is
\begin{equation}\label{eq:appendix_copula_state}
    \begin{split}
        \rho &= \mathcal{U}_A\otimes \mathcal{U}_B \rho_{in} \mathcal{U}_A^\dagger\otimes \mathcal{U}_B^\dagger\\
        &= \frac{1}{2^{n/2}}\sum_{i,j=0}^{2^{n/2}-1} \mathcal{U}_A|i\rangle_A \langle j|_A \mathcal{U}_A^\dagger \otimes \mathcal{U}_B|i\rangle_B \langle j|_B \mathcal{U}_B^\dagger
    \end{split}
\end{equation}
where $\mathcal{U}_A\otimes \mathcal{U}_B$ is the unitary operation realized by the the copula variational circuit. Note that we can write this circuit operation as separable since there are no entangling gates between registers $A$ and $B$ after the initial maximally entangled state preparation. See Fig.~\ref{fig:2D_discrete_copula_circuit} for reference.  
Since separable operations cannot change the amount of entanglement between the registers, and the initial state is maximally entangled, the marginal state $\rho_A$ must be proportional to the identity matrix. We see this by tracing out register $B$
\begin{equation}\label{eq:appendix_marginal_state}
    \begin{split}
        \rho_A &= \text{Tr}_B{(\rho)}\\
        &= \frac{1}{2^{n/2}}\sum_{i,j=0}^{2^{n/2}-1} \mathcal{U}_A|i\rangle_A \langle j|_A \mathcal{U}_A^\dagger\delta_{ij}\\
        &=\frac{1}{2^{n/2}}I_A,
    \end{split}
\end{equation}
where $I_A$ is the identity matrix on register $A$. 

We can also look at the state created by the copula circuit in terms of its representation in the computational basis. In general, this state is
\begin{equation}
\begin{split}
    |\psi\rangle &= \mathcal{U}_A\otimes \mathcal{U}_B |\psi_{in}\rangle\\
    &=\frac{1}{2^{n/4}}\sum_{i=0}^{2^{n/2}-1}\mathcal{U}_A|i\rangle_A\otimes \mathcal{U}_B|i\rangle_B\\
    &=\sum_{i,j=0}^{2^{n/2}-1}\sqrt{c_{ij}}|i\rangle_A\otimes|i\rangle_B
\end{split}    
\end{equation}
%\cjnote{I think the first line here should not have daggers}
where $c_{ij}$ are the probability amplitudes of the discrete copula which are learned by training the model, similar to Eq.~\eqref{eq:generator_quantum_state}. In this representation, the state the variational circuit prepares is
\begin{equation}
\begin{split}
     \rho &=|\psi\rangle \langle\psi| \\
     &=\sum_{i,j=0}^{2^{n/2}-1}\sum_{i',j'=0}^{2^{n/2}-1}\sqrt{c_{ij}c_{i'j'}}|i\rangle_A \langle i'|_A \otimes |j\rangle_B \langle j'|_B
\end{split}
\end{equation}
which is the same state given in Eq.~\eqref{eq:appendix_copula_state}. Again, we compute the partial trace with respect to register $B$ and obtain
\begin{equation}
\rho_A = \sum_{i,j=0}^{2^{n/2}-1}\sum_{i'=0}^{2^{n/2}-1}\sqrt{c_{ij}c_{i'j}} \:|i\rangle_A \langle i'|_A 
\end{equation}
where the last step is due to Eq.~\eqref{eq:appendix_marginal_state}.
When this state is measured in the computational basis $|u\rangle$, we obtain
\begin{equation}
    \begin{split}
        Pr(u) &= \text{Tr}(|u\rangle\langle u|\rho_A)\\
        &=\sum_{j=0}^{2^{n/2}-1}c_{uj}=\frac{1}{2^{n/2}}
    \end{split}
\end{equation}
which is the discrete uniform distribution. This last equation shows that the marginal distribution of the discrete copula $c_{ij}$ is uniformly distributed as promised.

\subsection{PIT and copula: analytic calculation for O data sets}\label{appendix:pit_calculation}
In this section, we briefly show how to compute the PIT analytically for one of the synthetic data sets used in the main text. We choose the O data set because it is the most informative, in our opinion. The same calculation can be carried out for any other data set.

The 2D O probability density can be written as
\begin{equation}
    P(x_1, x_2) = c \: \delta(1-x_1^2-x_2^2),
\end{equation}
where $c$ is a normalization constant.

The copula is given by Eq.~\eqref{eq:copula_analytical}
\begin{equation}\label{eq:appendix_2d_copula_transform}
\begin{split}
    & Pr(u_1,u_2) =\\
    & c \int dx_1\int dx_2 \delta(U_1-u_1)\delta(U_2-u_2)\delta(1-x_1^2-x_2^2),
\end{split}
\end{equation}
where $U_1=U_1(x_1)$ and $U_2=U_2(x_2)$ are given by Eq.~\eqref{eq:copula_marginals}

\begin{equation}\label{eq:appendix_O2D_U1}
\begin{split}
U_1 &= c\int_{-\infty}^{x_1}dx_1' \int_{-\infty}^{\infty}dx'_2 \delta(1-x_1'^2-x_2'^2)\\
&=c\int_{-\infty}^{x_1}dx_1' \int_{-\infty}^{\infty}dx'_2 \frac{\delta(x_2'-\sqrt{1-x_1'^2})+ \delta(x_2'+\sqrt{1-x_1'^2})}{2|x_2'|}\\
&=c\int_{-1}^{x_1} \frac{dx_1'}{\sqrt{1-x_1'^2}} \\
&=\frac{1}{\pi}\arcsin{(x_1)}+\frac{1}{2},
\end{split}
\end{equation}
where we have use the fact that
\begin{equation}\label{eq:appendix_delta_f(x)}
    \delta(f(\chi)) = \sum_i\frac{\delta(\chi-\chi_i)}{|f'(\chi_i)|},
\end{equation}
where $\chi_i$ are the roots of $f(\chi)$ and $f'(\chi)$ is its derivative. In the last step, we give the exact normalization for the cumulative marginals.

By similar arguments, we obtain
\begin{equation}\label{eq:appendix_O2D_U2}
U_2 =\frac{1}{\pi}\arcsin{(x_2)}+\frac{1}{2}.
\end{equation}

With these cumulative marginals, we can compute the copula explicitly by calculating the integral in Eq.~\eqref{eq:appendix_2d_copula_transform} as follows
\begin{equation}
\begin{split}
    & Pr(u_1,u_2) =\\
    &\frac{1}{\pi} \int dx_1\int dx_2 \delta(\frac{1}{\pi}\arcsin{(x_1)}+\frac{1}{2}-u_1)\\
    &\delta(\frac{1}{\pi}\arcsin{(x_2)}+\frac{1}{2}-u_2) \delta(1-x_1^2-x_2^2).
\end{split}
\end{equation}
Making use of Eqs.~\eqref{eq:appendix_O2D_U1}, \eqref{eq:appendix_O2D_U2}, and \eqref{eq:appendix_delta_f(x)} we can write
\begin{equation}
\begin{split}
    & Pr(u_1,u_2) =\\
    &\pi \int dx_1\int dx_2 \frac{\delta(x_1-\sin{(\pi(u_1-\frac{1}{2}))})}{\sqrt{1-\sin^2(\pi(u_1-\frac{1}{2}))}}\cdot\\
    &\frac{\delta(x_2-\sin{(\pi(u_2-\frac{1}{2}))})}{\sqrt{1-\sin^2(\pi(u_2-\frac{1}{2}))}} \delta(1-x_1^2-x_2^2).
\end{split}
\end{equation}

Computing both integrals, we obtain
\begin{equation}
\begin{split}
    & Pr(u_1,u_2) =\\
    &\pi \frac{\delta(1-\sin^2{(\pi(u_1-\frac{1}{2}))} - \sin^2{(\pi(u_2-\frac{1}{2}))})}{|\cos{(\pi(u_1-\frac{1}{2}))}||\cos{(\pi(u_2-\frac{1}{2}))}|}.
\end{split}
\end{equation}
The last expression is non-zero only when
\begin{equation}
    1-\sin^2{(\pi(u_1-\frac{1}{2}))} - \sin^2{(\pi(u_2-\frac{1}{2}))} = 0
\end{equation}
or equivalently, when
\begin{equation}
    |\cos{(\pi(u_1-\frac{1}{2}))}| =  |\sin{(\pi(u_2-\frac{1}{2}))}|,
\end{equation}
which happens when
\begin{equation}
    u_2 = \left\{\begin{array}{cc}
        u_1+\frac{1}{2} & 0\leq u_1 < \frac{1}{2} \\
        -u_1+\frac{1}{2} & 0\leq u_1 < \frac{1}{2} \\
        u_1-\frac{1}{2} & \frac{1}{2} \leq u_1 \leq 1\\
        -u_1+\frac{3}{2} & \frac{1}{2} \leq u_1 \leq 1
    \end{array} \right.
\end{equation}
in accordance with Fig.~\ref{fig:data_visulaization_2d}.

A similar analysis for the O 3D data set, with $Pr(x_1,x_2,x_3) = c\delta(1-x_1^2-x_2^2-x_3^2)$ yields constant marginals, and therefore linear cumulative marginals, $U_i(x_i) = \frac{1}{2}(x_i+1)$. Thus $P(u_1,u_2,u_3)\propto \delta(\frac{1}{4}-(u_1-\frac{1}{2})^2-(u_2-\frac{1}{2})^2-(u_3-\frac{1}{2})^2)$, which is the uniform distribution on the surface of a sphere of radius $\frac{1}{2}$ centered at $(\frac{1}{2},\frac{1}{2},\frac{1}{2})$ as shown in Fig.~\ref{fig:data_visulaization_3d}. 
%\tenote{if we want to include that - or figure ~\ref{fig:data_visulaization_3d_O_20}}

\subsection{Resource summary} \label{appendix:resource_summary} 
\begin{figure}[t]
    \centering
       \includegraphics[scale=0.5]{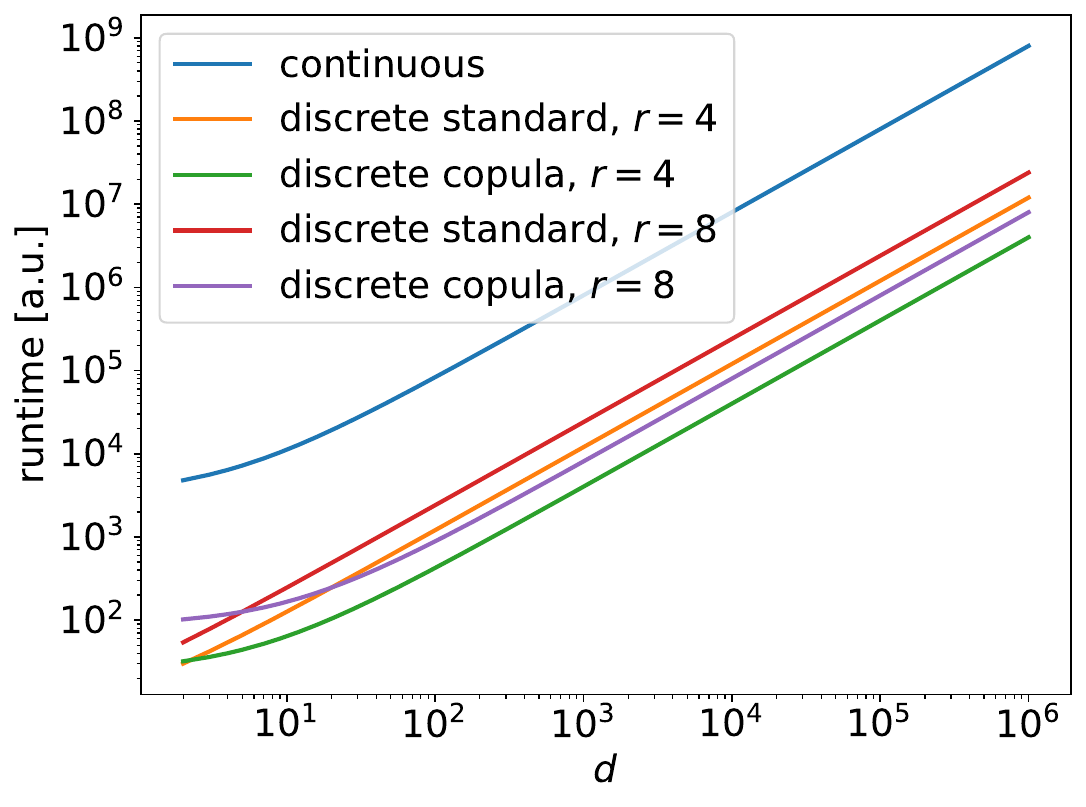}
        \caption{Estimated runtime per generated sample on quantum hardware as a function of the dimension of the data. We have used the values shown in Table \ref{tab:resource_comparison}, for fixed discretization grid ($r=4$ and $r=8$) and 3 and 8 variational circuit blocks for the discrete and continuous architectures, respectively. For the continuous architecture 100 shots are shown. The discrete architecture requires only 1 shot per sample. The asymptotic behavior is linear in $d$ with a constant separation between the continuous and discrete architectures given by the number of shots, $N_s$, required to compute the output of the continuous architecture. The discrete copula architecture is more efficient than the others mainly due to the fact that the cost per cicuit block does not directly on the dimension of the data.  
        }\label{fig:runtime} 
\end{figure}

In this section, we calculate the number of gates and  the depth of the all quantum circuits used in this work. With this information, we also estimate the scaling of the runtime required to generate a single data sample from such circuits in a quantum computer. The results are presented in Table \ref{tab:resource_comparison}. The runtime scales linearly with the dimension of the data set for all cases, this can be seen in Fig.~\ref{fig:runtime}. 
This figure also shows that even as the data dimension $d$ increases the continuous circuit run time is always higher than the discrete circuits, this  due the continuous circuit needing more shots per sample. 
 An interesting fact is that, asymptotically,  the continuous and discrete architectures can run at similar rates if the discretization parameter $r>266$ for a fixed number of circuit blocks. This is much higher than $r=32$, which would be similar to single precision operations in classical hardware.

\begin{table*}
\caption{Resource estimation per generated sample and comparison among all quantum circuits used in this work for $d$-dimensional data. Note that $r$ is the number of qubits used to discretize each dimension of the data and it only plays a role in the discrete architectures. Here, $N_b$ is the number of the repeated blocks of quantum circuit and $N_s$ is the number of shots required to generate a single data sample. We estimate that $N_s>100$ shots will be required to estimate the expectation value given in \eqref{eq:continuous_output_samples} with a reasonable variance. Note that the discrete architectures require only 1 shot per data sample.  Also, note that the copula circuit requires an initializing (Init). in the table) operation which only occurs once and therefore, we account it separately. %\alnote{Could we consider also the resources per iteration? For Discrete we sample like 10,000 images per iteration, while for continous we need to do the measurements but can then fairly easily compute the synthetic images?} \crnote{If we want to find the runtime per iteration, we need to multiply the runtime below by the number of calls of the circuit during one iteration. That number will be a function of the batch size (or number of samples) and the number of free parameters (times 2) for the continuous and a function of $d$, $r$ (number of free parameters) and batch size (or number of samples) for the discrete QGANs. All comes from evaluating the gradient. For the QCBM the factor will be a function of the population size of CMA-ES for the continuous, and for the discrete it'll be a function of $d$, $r$ (number of free parameters) and the population size in CMA-ES. I think it will quickly become very complicated and also dependent on non-quantum-circuit hyper-parameters. Currently, all numbers below depend only on the quantum circuits. Regarding you second point. the continuous requires many more samples to generate the synthetic data in a real quantum device. In our simulations, this is not the case, but that is not relevant for this table.} 
\label{tab:resource_comparison}
}
\centering
\begin{tabular}[t]{lccccccc}
\toprule

Circuit & qubits & Init. gates & Init. depth& gates/block  & depth/block & Runtime/sample \\
\midrule
\cellcolor{gray!6}{Continuous} & \cellcolor{gray!6}{$d$} & \cellcolor{gray!6}{---} & \cellcolor{gray!6}{---} & \cellcolor{gray!6}{$5d$} & \cellcolor{gray!6}{$d+4$} & \cellcolor{gray!6}{$N_sN_b(d+4)$} \\
Discrete standard & $rd$ & --- & --- & $3rd-2$ & $rd+2$ & $N_b(rd+2)$\\
\cellcolor{gray!6}{Discrete copula} & \cellcolor{gray!6}{$rd$} & \cellcolor{gray!6}{$rd$} & \cellcolor{gray!6}{$r(d-1)+1$} & \cellcolor{gray!6}{$\frac{rd}{2}(r+5)$} & \cellcolor{gray!6}{$\frac{r(r-1)}{2}+3$} & \cellcolor{gray!6}{$r(d-1)+1 + N_b(\frac{r(r-1)}{2}+3)$}\\

\bottomrule
\end{tabular}
\end{table*}

\subsection{Training behavior}\label{appendix:training_curves} For completeness, we present the learning curves for a representative subset of the experiments we ran in this work. We show the behavior of the Kullback-Leibler divergence as a function of training epoch, for all cases. Note that for the QCBM training, these curves represent directly the loss function used during training whereas for the training of the QGANs, they do not. In that case, the losses are the cross entropy terms given in Eqs.~\eqref{eq:discriminator_loss} and \eqref{eq:generator_loss}, in the main text. However, we find more informative to present the training evolution of our main performance metric.

\begin{figure*}[t]
\centering
\includegraphics[width=0.49\textwidth]{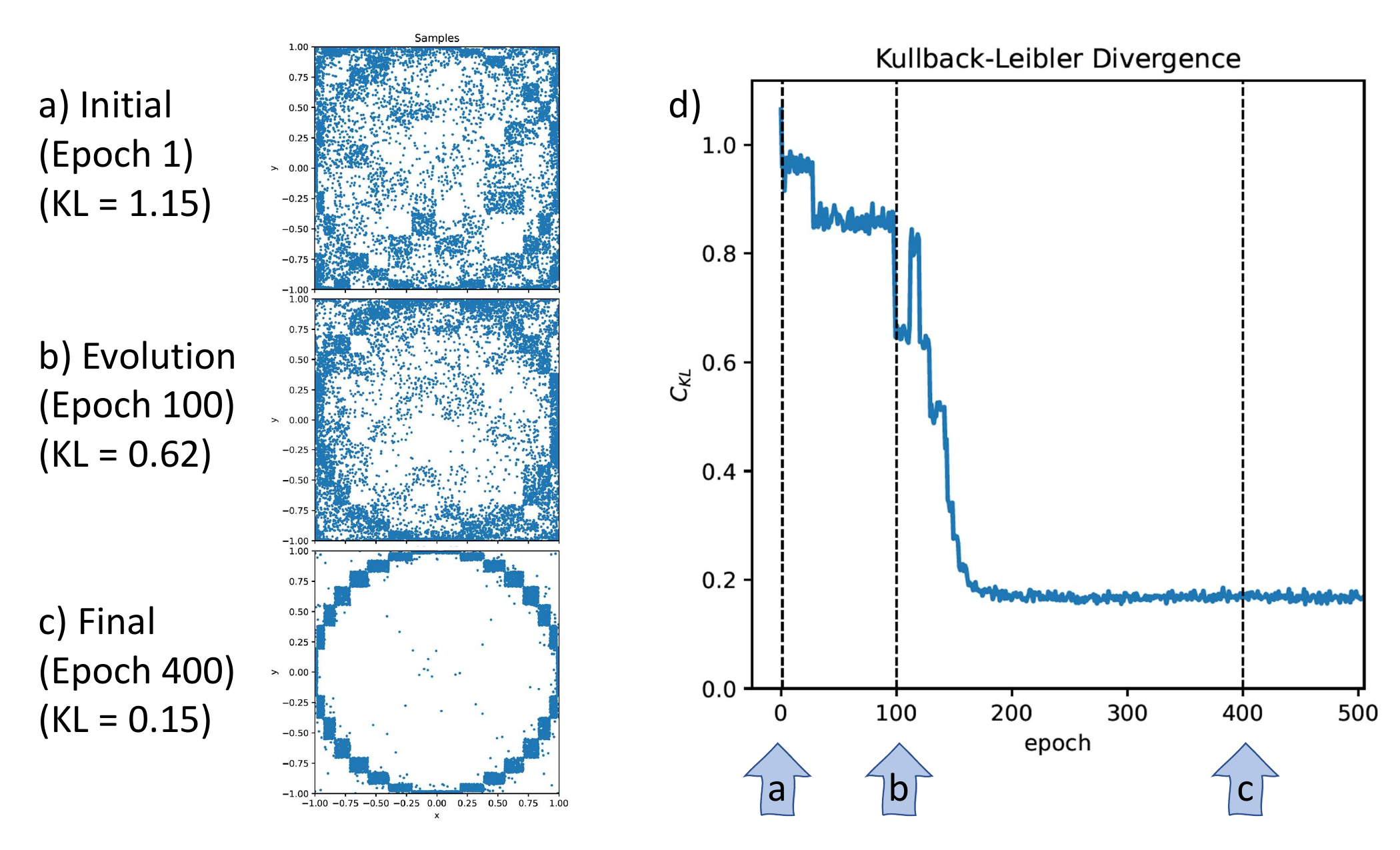}
\includegraphics[width=0.49\textwidth]{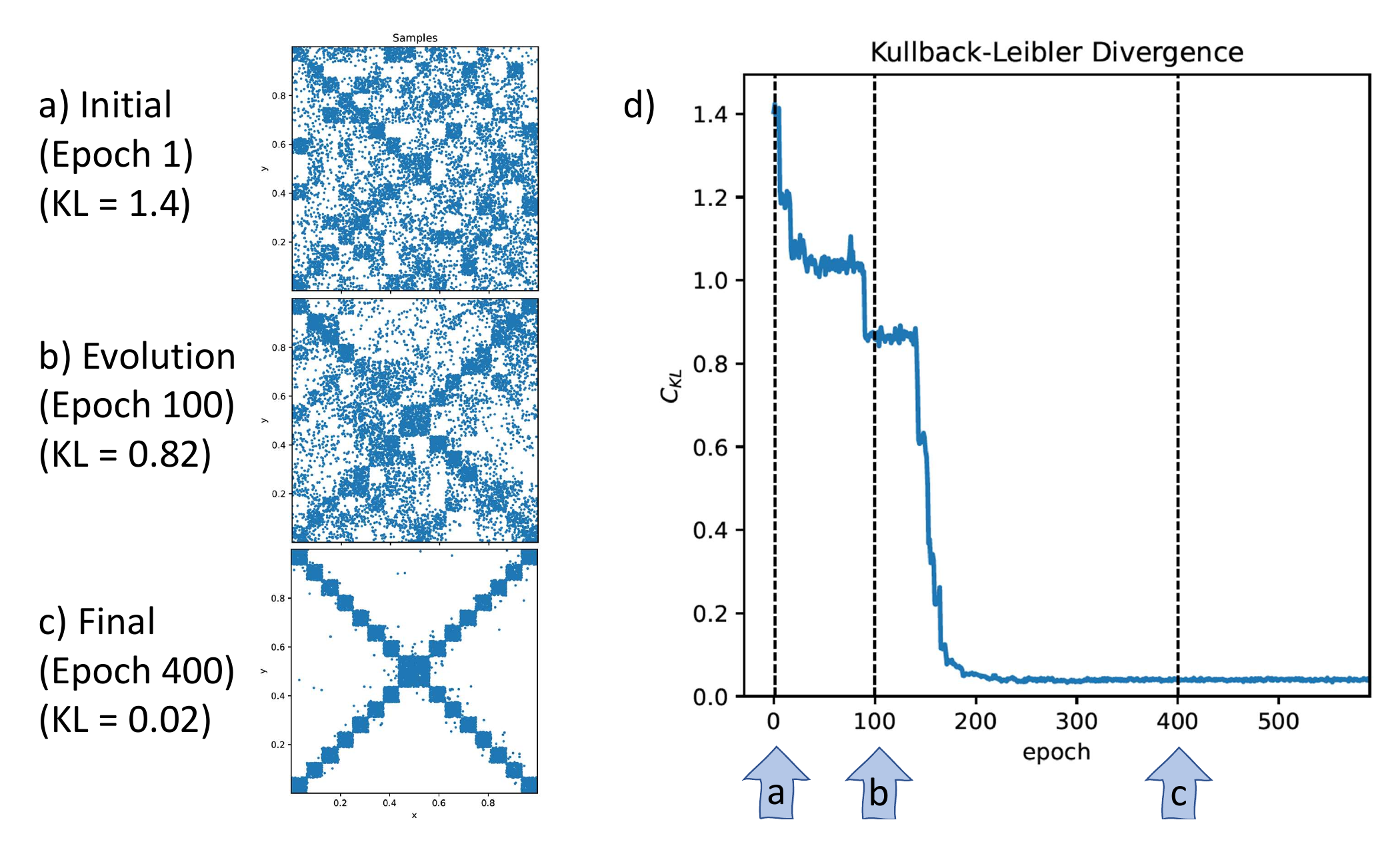}
\caption{Example of generated data using our trained quantum models for two of the data sets used in this work: O 2D (left) and X 2D (right), both trained via QCBM with one circuit block of the discrete copula architecture and PIT data, at different phases of the training and the associated KL over the training epochs. a) Generated samples at beginning of training (Epoch 1), b) during training (Epoch 100), c) at the end (Epoch 400), d) corresponding KL evolution over Epochs. Note the discretization errors in both examples, which are a direct result of adding external noise to the discrete sampled values. See the main text for details. } \label{fig:KL_evolution}
\end{figure*}

\begin{figure*}[t]
\centering
\includegraphics[width=0.49\textwidth]{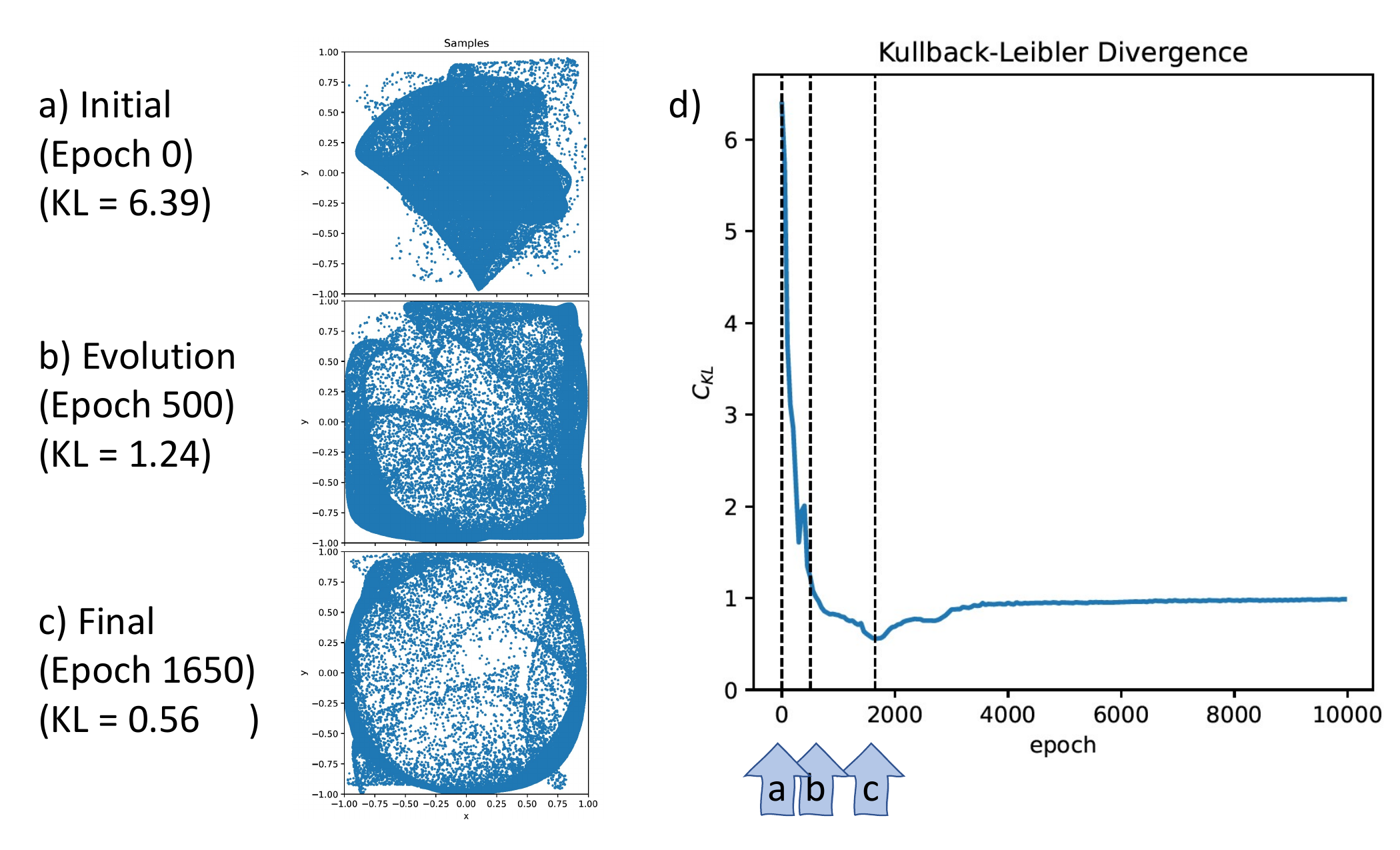}
\includegraphics[width=0.49\textwidth]{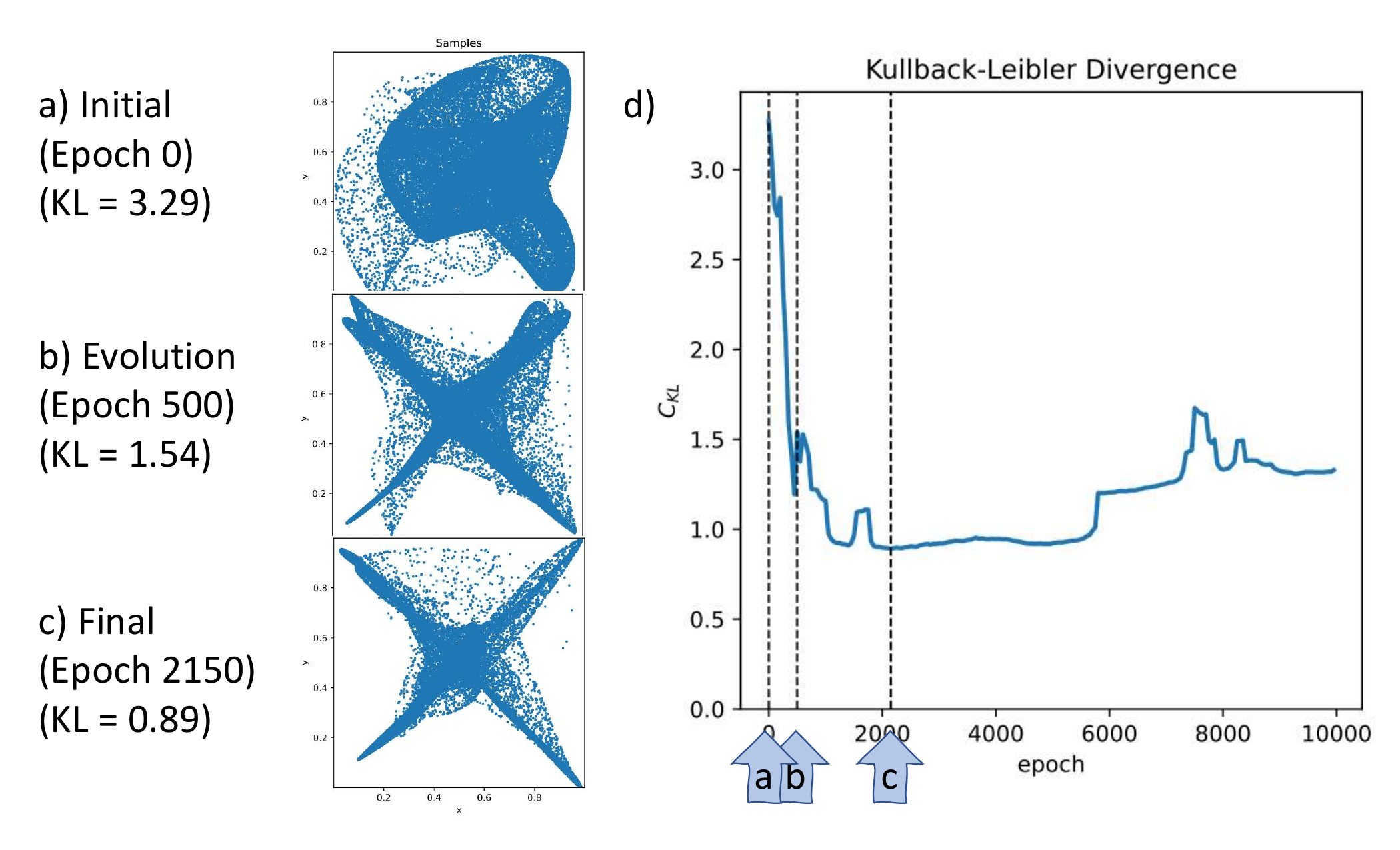} 
\caption{Example of generated data using our trained quantum models for two of the data sets used in this work: O 2D (left) and X 2D (right), both trained via continuous architecture, at different phases of the training and the associated KL over the training epochs. a) Generated samples at beginning of training (Epoch 0), b) during training (Epoch 500), c) at the end (Epoch 1650, (left) and Epoch 2150 (right)), d) corresponding KL evolution over Epochs.} \label{fig:KL_evolution_continuous}
\end{figure*}

In Fig.~\ref{fig:KL_evolution}, we show two examples of the training evolution of the discrete copula architecture trained via QCBM. In addition to the training curves, we show examples of the synthetic data sampled from our simulated quantum generative model. The correlation between the value of the KL divergence and the resemblance of the samples is clear. Note the ``boxy" nature of the samples. This is a result of the large discretization error, which is unavoidable in relatively small quantum devices. This is in contrast with the behavior of the continuous quantum architecture, which we present in Fig.\ref{fig:KL_evolution_continuous}. There, we show two training examples of the QCBM-trained continuous architecture. As before, the intuitive correlation between the value of the KL divergence and the quality of the samples is clear. Also, note that there is no discretization error, due to the fact that the generated samples are continuous valued.

\begin{figure*}
     \centering
     \begin{subfigure}[b]{0.45\textwidth}
         \centering
         \caption{\textbf{Mixed Gaussian}}
         \includegraphics[width=\textwidth]{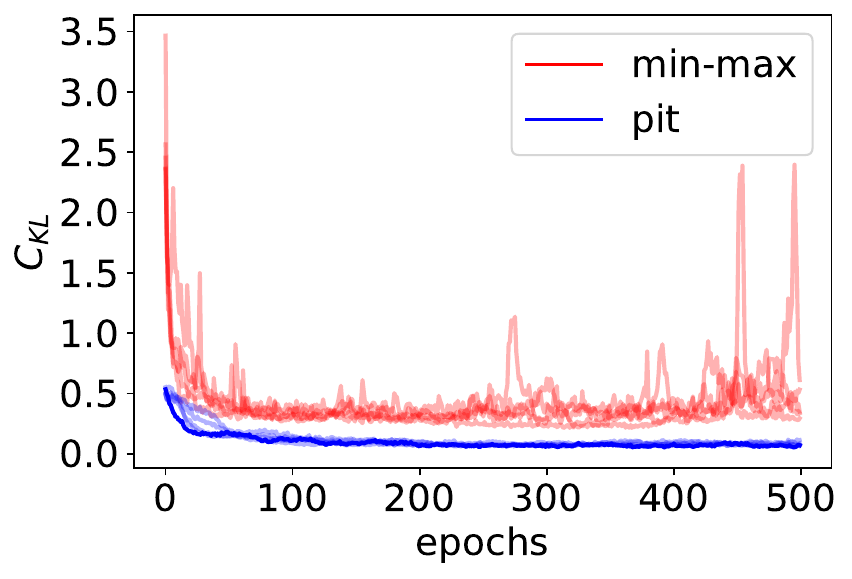}
         
     \end{subfigure}
     %\hfill
     \begin{subfigure}[b]{0.45\textwidth}
         \centering
         \caption{\textbf{X}}
         \includegraphics[width=\textwidth]{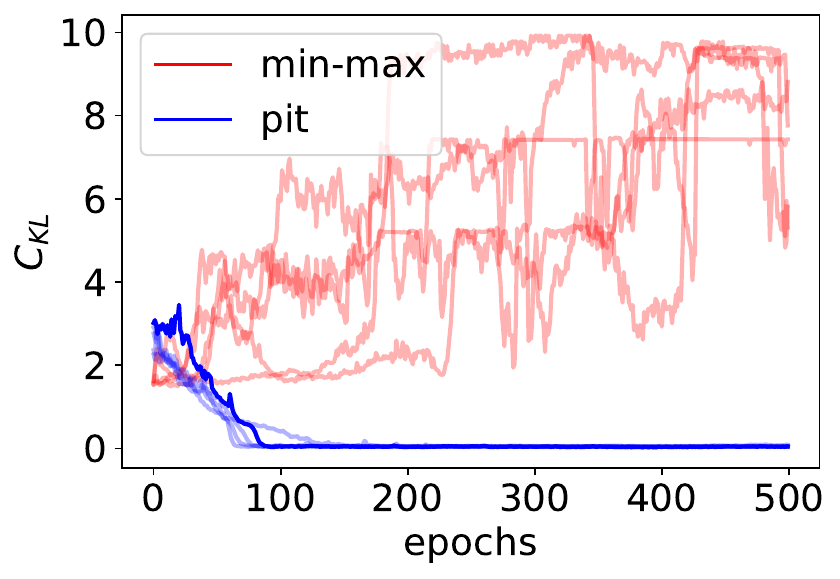}
         
     \end{subfigure}
     %\hfill
     \begin{subfigure}[b]{0.45\textwidth}
         \centering
         \caption{\textbf{O}}
         \includegraphics[width=\textwidth]{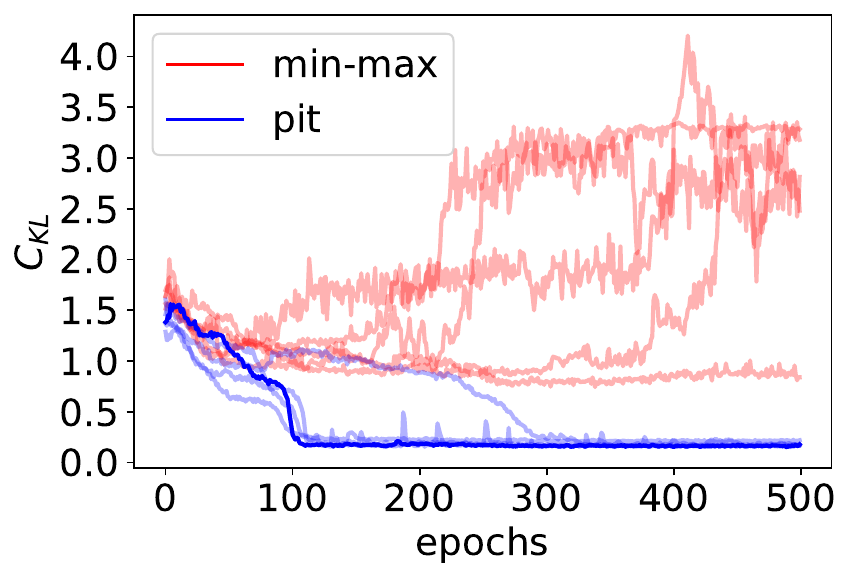}
         
     \end{subfigure}
     %\hfill
     \begin{subfigure}[b]{0.45\textwidth}
         \centering
         \caption{\textbf{Stocks}}
         \includegraphics[width=\textwidth]{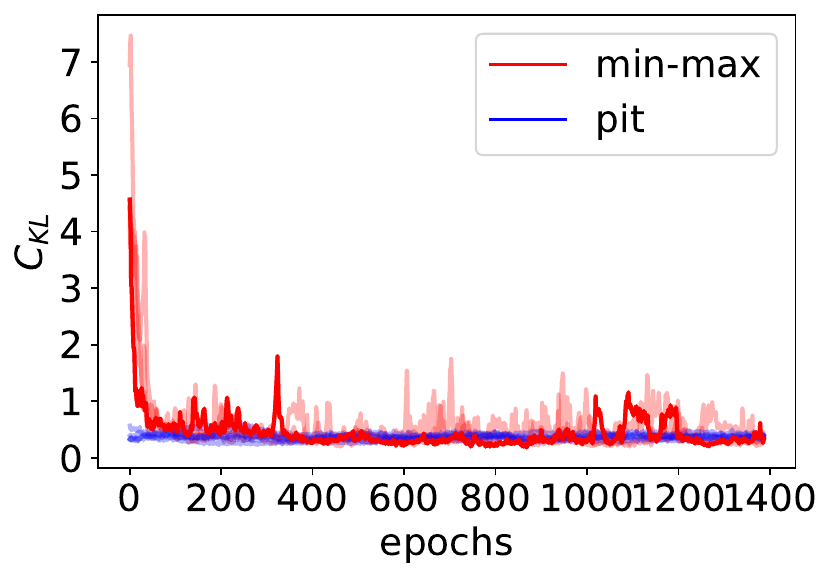}
         
     \end{subfigure}
        \caption{KL divergence as a function of training epoch for the QGAN discrete architectures and all 2D data sets. Here we contrast the training between the 2 data transformations we used in this work, PIT and Min-Max normalization. We chose 50 gradient evaluations per epoch (or a mini batch size of 1000) for all data sets except the stocks, for which we used 18 (or a mini batch size of 104).}
        \label{fig:discrete_gan_learning_curves_2D}
\end{figure*}

\begin{figure*}
     \centering
     \begin{subfigure}[b]{0.45\textwidth}
         \centering
         \caption{\textbf{Mixed Gaussian}}
         \includegraphics[width=\textwidth]{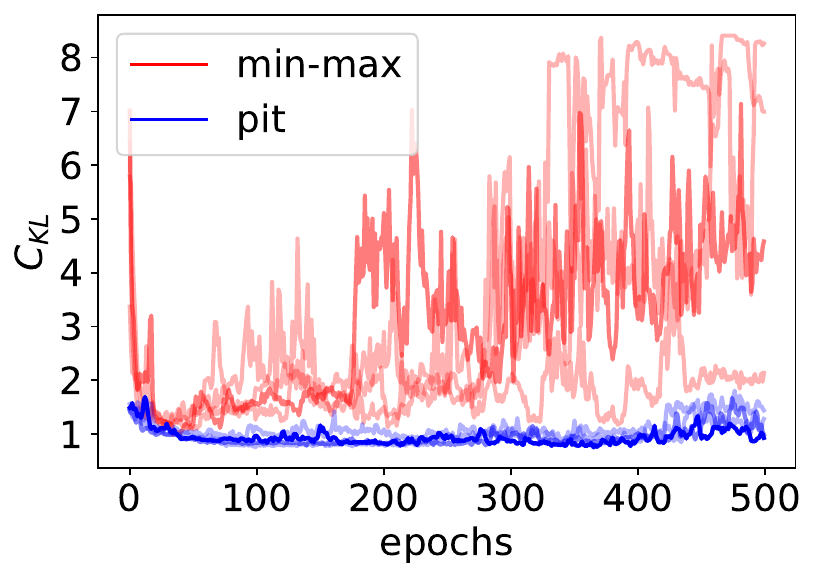}
         
     \end{subfigure}
     %\hfill
     \begin{subfigure}[b]{0.45\textwidth}
         \centering
         \caption{\textbf{X}}
         \includegraphics[width=\textwidth]{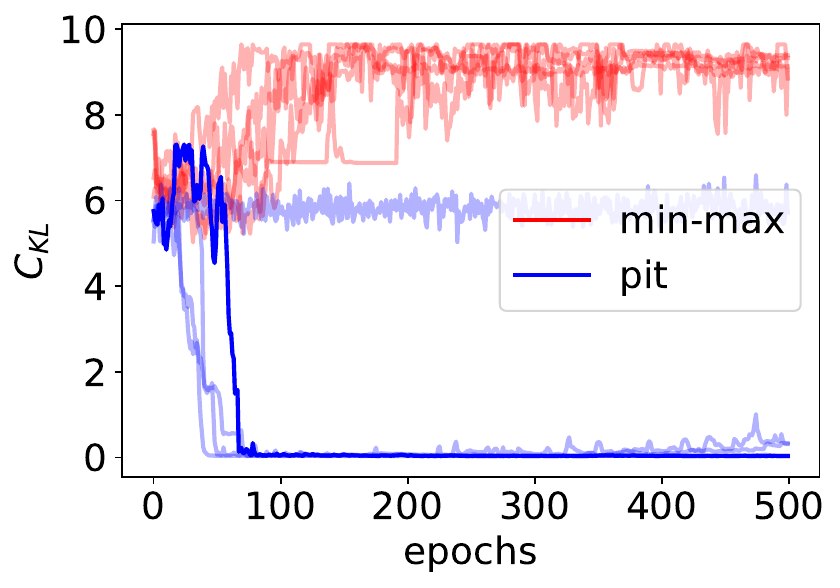}
         
     \end{subfigure}
     %\hfill
     \begin{subfigure}[b]{0.45\textwidth}
         \centering
         \caption{\textbf{O}}
         \includegraphics[width=\textwidth]{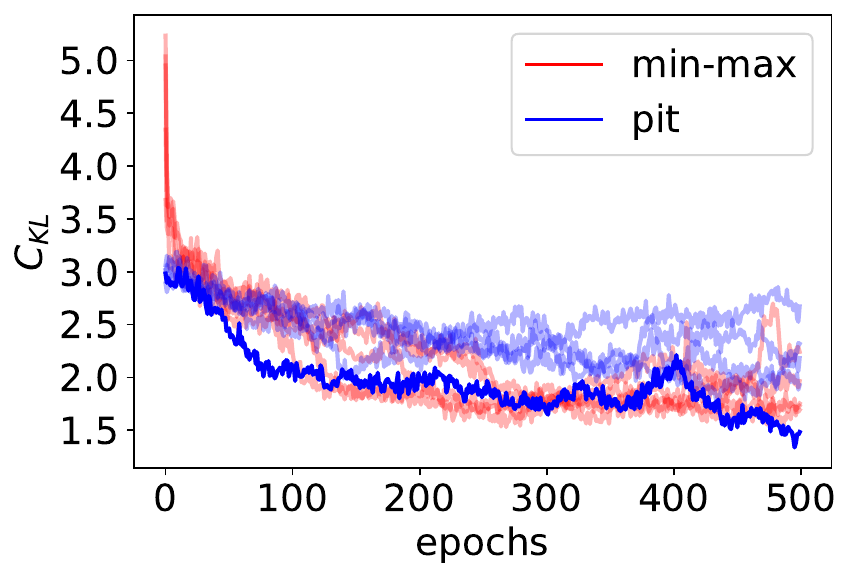}
         
     \end{subfigure}
     %\hfill
     \begin{subfigure}[b]{0.45\textwidth}
         \centering
         \caption{\textbf{Stocks}}
         
         \includegraphics[width=\textwidth]{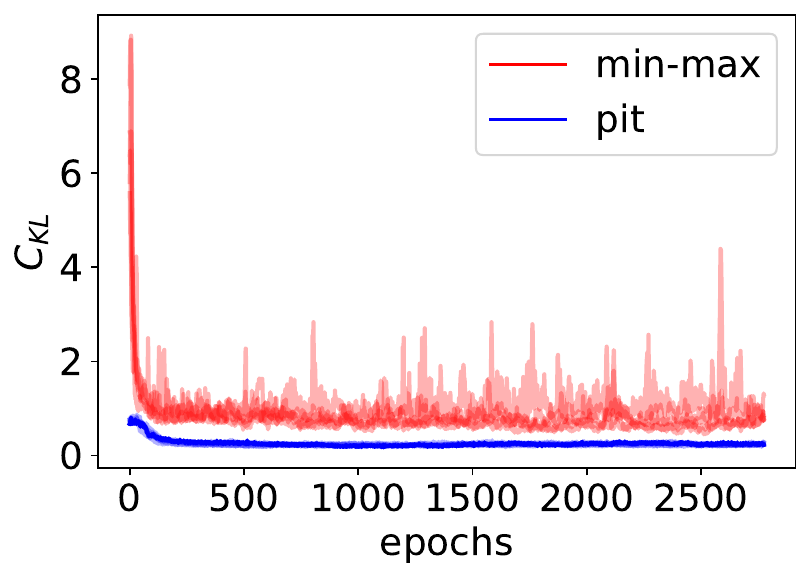}
         
     \end{subfigure}
        \caption{KL divergence as a function of training epoch for the QGAN discrete architectures and all 3D data sets. Here we contrast the training between the 2 data transformations we used in this work, PIT and Min-Max normalization. We chose 100 gradient evaluations per epoch (or a mini batch size of 1000) for all data sets except the stocks, for which we used 18 (or a mini batch size of 104).}        \label{fig:discrete_gan_learning_curves_3D}
\end{figure*}

\begin{figure*}
     \centering
     \begin{subfigure}[b]{0.45\textwidth}
         \centering
         \caption{\textbf{Mixed Gaussian}}
         \includegraphics[width=\textwidth]{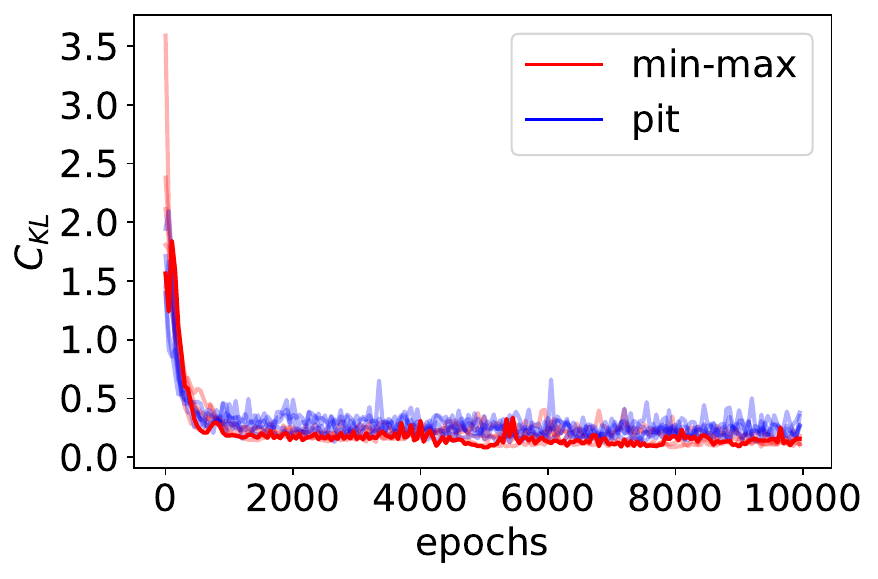}
     \end{subfigure}
     \begin{subfigure}[b]{0.45\textwidth}
         \centering
         \caption{\textbf{X}}
         \includegraphics[width=\textwidth]{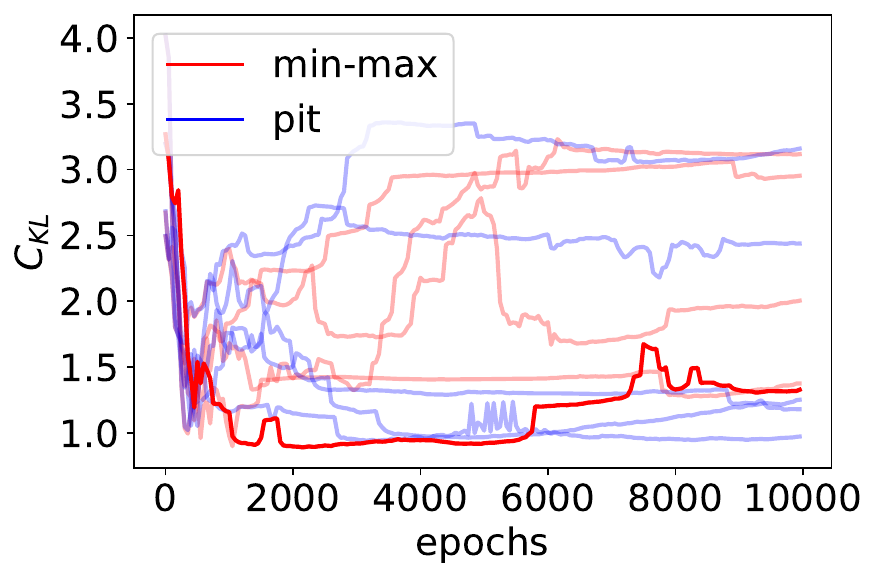}
     \end{subfigure}
     %\hfill
     \begin{subfigure}[b]{0.45\textwidth}
         \centering
         \caption{\textbf{O}}
         \includegraphics[width=\textwidth]{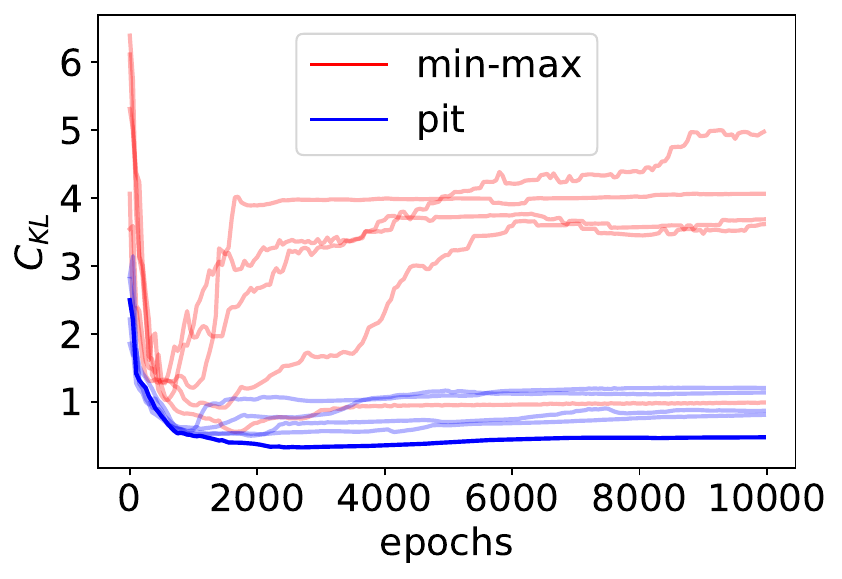}
     \end{subfigure}
     %\hfill
     \begin{subfigure}[b]{0.45\textwidth}
         \centering
         \caption{\textbf{Stocks}}
         \includegraphics[width=\textwidth]{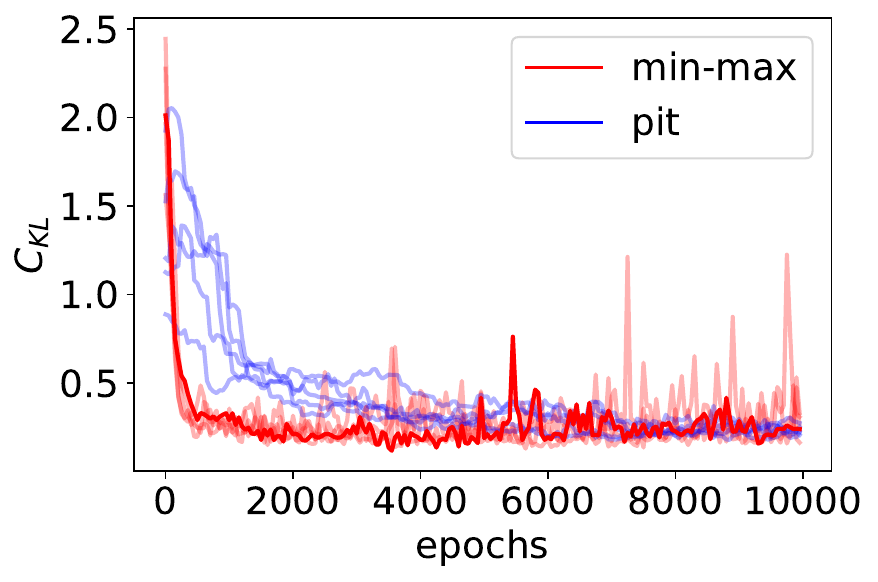}
     \end{subfigure}
        \caption{KL divergence as a function of training epoch for the QGAN continuous architecture and all 2D data sets. Here we contrast the training between the 2 data transformations we used in this work, PIT and Min-Max normalization. We chose to train the 2D experiments with 1 gradient computation per epoch (or a mini batch size of 50000).}
        \label{fig:continuous_gan_learning_curves_2D}
\end{figure*}

\begin{figure*}
     \centering
     \begin{subfigure}[b]{0.45\textwidth}
         \centering
         \caption{\textbf{Mixed Gaussian}}
         \includegraphics[width=\textwidth]{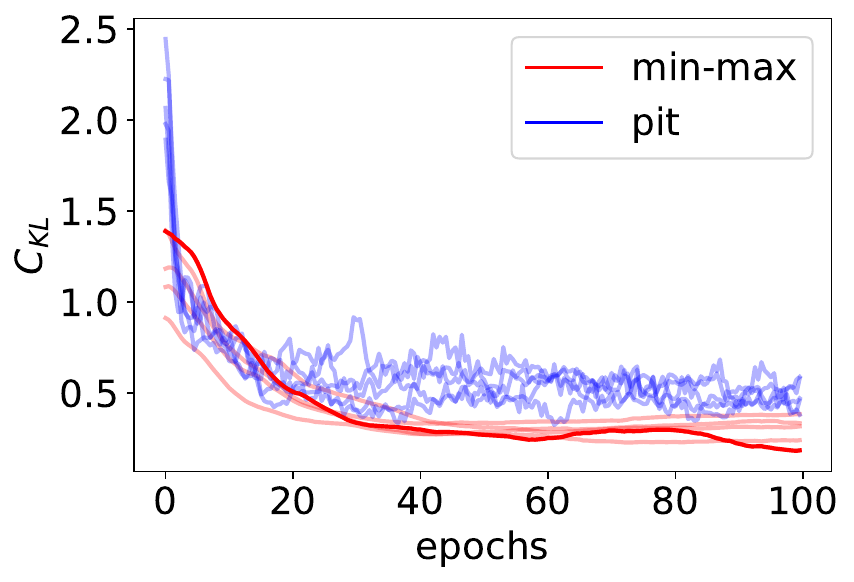}
         
     \end{subfigure}
     %\hfill
     \begin{subfigure}[b]{0.45\textwidth}
         \centering
         \caption{\textbf{X}}
         \includegraphics[width=\textwidth]{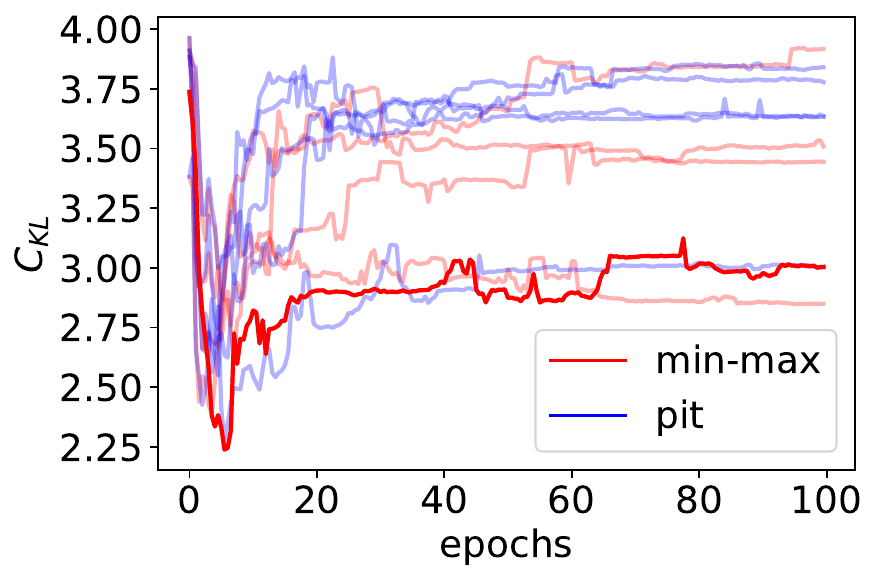}
         
     \end{subfigure}
     %\hfill
     \begin{subfigure}[b]{0.45\textwidth}
         \centering
         \caption{\textbf{O}}
         \includegraphics[width=\textwidth]{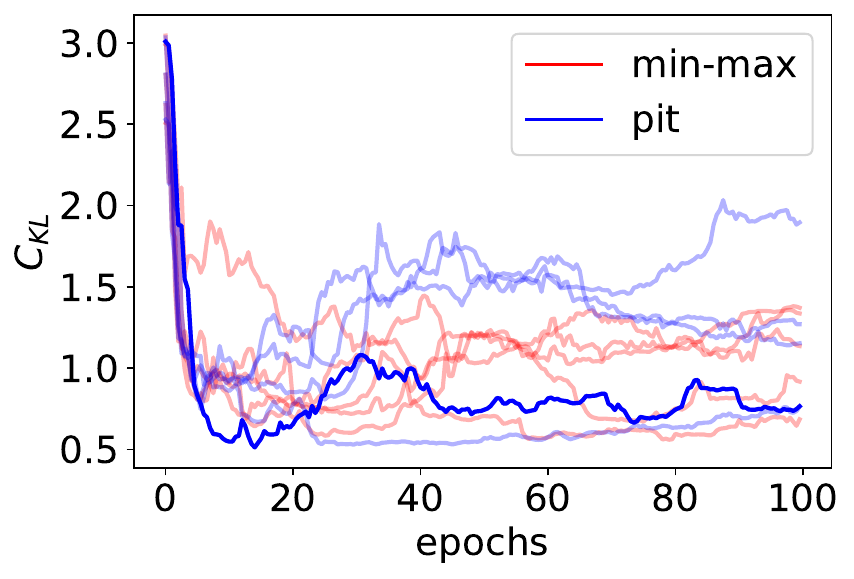}
         
     \end{subfigure}
     %\hfill
     \begin{subfigure}[b]{0.45\textwidth}
         \centering
         \caption{\textbf{Stocks}}
         \includegraphics[width=\textwidth]{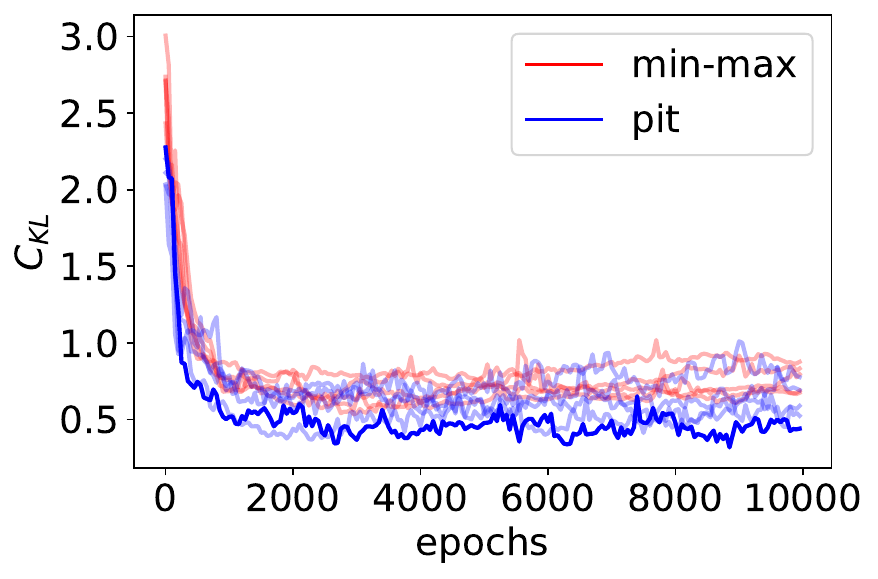}
         
     \end{subfigure}
        \caption{KL divergence as a function of training epoch for the QGAN continuous architecture for all the 3D data sets. Here we contrast the training between the 2 data transformations we used in this work, PIT and Min-Max normalization. We chose to use 100 gradient computations per epoch (or a mini batch size of 1000) for all data sets except the stocks, for which we used 1 gradient evaluation (or a batch size of 1877).}
        \label{fig:continuous_gan_learning_curves_3D}
\end{figure*}

In Figs.~\ref{fig:discrete_gan_learning_curves_2D} and \ref{fig:discrete_gan_learning_curves_3D}, we present the training curves for all QGAN experiments we ran using the discrete architectures. Note that for most of the cases, training using the copula architecture and the PIT is more stable and with less sudden jumps. Finally, in Figs.~\ref{fig:continuous_gan_learning_curves_2D} and \ref{fig:continuous_gan_learning_curves_3D}, we present the results for the training using the continuous architecture described in the main text. Here, there is no clear advantage regarding the data transformations. In fact, since there is a unique continuous architecture tested in this work, no clear difference in training performance can be seen between the 2 data transformations, min-max and PIT.

\end{document}